\pdfoutput=1
\documentclass[12pt,a4paper]{article}

\usepackage{ifthen} 
\newboolean{pdflatex}
\setboolean{pdflatex}{true} 
\usepackage{multirow}

\newboolean{articletitles}
\setboolean{articletitles}{true} 

\newboolean{uprightparticles}
\setboolean{uprightparticles}{false} 

\def\paperauthors{LHCb collaboration} 
\def\paperasciititle{Observation of enhanced double parton scattering in proton-lead collisions at nucleon nucleon center-of-mass enregy of 8.16 TeV}
\def\papertitle{Observation of enhanced double parton scattering in proton-lead collisions at $\sqsnn=8.16\tev$ }
\def\paperkeywords{{High Energy Physics}, {LHCb}} 
\def\papercopyright{\the\year\ CERN for the benefit of the LHCb collaboration} 
\def\paperlicence{CC BY 4.0 licence}
\def\paperlicenceurl{https://creativecommons.org/licenses/by/4.0/}

\usepackage[top=1in, bottom=1.25in, left=1in, right=1in]{geometry}

\usepackage{microtype}
\usepackage{lineno}  
\usepackage{xspace} 
\usepackage{caption} 

\usepackage{graphicx}  
\usepackage{color}
\usepackage{colortbl}
\graphicspath{{./figs/}} 

\usepackage{amsmath} 
\usepackage{amssymb}
\usepackage{amsfonts}
\usepackage{upgreek} 

\newcommand*\patchAmsMathEnvironmentForLineno[1]{%
\expandafter\let\csname old#1\expandafter\endcsname\csname #1\endcsname
\expandafter\let\csname oldend#1\expandafter\endcsname\csname
end#1\endcsname
 \renewenvironment{#1}%
   {\linenomath\csname old#1\endcsname}%
   {\csname oldend#1\endcsname\endlinenomath}%
}
\newcommand*\patchBothAmsMathEnvironmentsForLineno[1]{%
  \patchAmsMathEnvironmentForLineno{#1}%
  \patchAmsMathEnvironmentForLineno{#1*}%
}
\AtBeginDocument{%
\patchBothAmsMathEnvironmentsForLineno{equation}%
\patchBothAmsMathEnvironmentsForLineno{align}%
\patchBothAmsMathEnvironmentsForLineno{flalign}%
\patchBothAmsMathEnvironmentsForLineno{alignat}%
\patchBothAmsMathEnvironmentsForLineno{gather}%
\patchBothAmsMathEnvironmentsForLineno{multline}%
\patchBothAmsMathEnvironmentsForLineno{eqnarray}%
}

\usepackage{hyperxmp}

\usepackage[pdftex,
            pdfauthor={\paperauthors},
            pdftitle={\paperasciititle},
            pdfkeywords={\paperkeywords},
            pdfcopyright={Copyright (C) \papercopyright},
            pdflicenseurl={\paperlicenceurl}]{hyperref}

\usepackage[colorinlistoftodos,textsize=scriptsize]{todonotes}

\usepackage[bottom,flushmargin,hang,multiple]{footmisc}

\usepackage[all]{hypcap} 

\usepackage{xspace} 
\usepackage{upgreek}

\def\lhcb   {\mbox{LHCb}\xspace}

\def\lhc    {\mbox{LHC}\xspace}

\def\tevatron {Tevatron\xspace}

\def\MagUp {\mbox{\em Mag\kern -0.05em Up}\xspace}

\ifthenelse{\boolean{uprightparticles}}%
{

 \def\Pmu         {\ensuremath{\upmu}\xspace}

 \def\Ppi         {\ensuremath{\uppi}\xspace}

 \def\Ppsi        {\ensuremath{\uppsi}\xspace}

 \def\PDelta      {\ensuremath{\Delta}\xspace}                 
 \def\PXi         {\ensuremath{\Xi}\xspace}                 
 \def\PLambda     {\ensuremath{\Lambda}\xspace}                 
 \def\PSigma      {\ensuremath{\Sigma}\xspace}                 
 \def\POmega      {\ensuremath{\Omega}\xspace}                 
 \def\PUpsilon    {\ensuremath{\Upsilon}\xspace}

 \def\PB      {\ensuremath{\mathrm{B}}\xspace}                 
                  
 \def\PD      {\ensuremath{\mathrm{D}}\xspace}

 \def\PJ      {\ensuremath{\mathrm{J}}\xspace}                 
 \def\PK      {\ensuremath{\mathrm{K}}\xspace}

 \def\Pc      {\ensuremath{\mathrm{c}}\xspace}

 \def\Pi      {\ensuremath{\mathrm{i}}\xspace}

 \def\Ps      {\ensuremath{\mathrm{s}}\xspace}

 \def\thebaroffset{0.0em}
}
{

 \def\Pmu         {\ensuremath{\mu}\xspace}

 \def\Ppi         {\ensuremath{\pi}\xspace}

 \def\Ppsi        {\ensuremath{\psi}\xspace}                 
                  
 \mathchardef\PDelta="7101
 \mathchardef\PXi="7104
 \mathchardef\PLambda="7103
 \mathchardef\PSigma="7106
 \mathchardef\POmega="710A
 \mathchardef\PUpsilon="7107
                  
 \def\PB      {\ensuremath{B}\xspace}                 
                  
 \def\PD      {\ensuremath{D}\xspace}

 \def\PJ      {\ensuremath{J}\xspace}                 
 \def\PK      {\ensuremath{K}\xspace}

 \def\Pc      {\ensuremath{c}\xspace}

 \def\Pi      {\ensuremath{i}\xspace}

 \def\Ps      {\ensuremath{s}\xspace}

 \def\thebaroffset{0.18em}
}
\newcommand{\offsetoverline}[2][\thebaroffset]{\kern #1\overline{\kern -#1 #2}}%

\makeatletter
\ifcase \@ptsize \relax
  \newcommand{\miniscule}{\@setfontsize\miniscule{4}{5}}
\or
  \newcommand{\miniscule}{\@setfontsize\miniscule{5}{6}}
\or
  \newcommand{\miniscule}{\@setfontsize\miniscule{5}{6}}
\fi
\makeatother

\DeclareRobustCommand{\optbar}[1]{\shortstack{{\miniscule (\rule[.5ex]{1.25em}{.18mm})}
  \\ [-.7ex] $#1$}}

\def\mun        {{\ensuremath{\Pmu^-}}\xspace} 

\def\mumu       {{\ensuremath{\Pmu^+\Pmu^-}}\xspace}

\def\squark    {{\ensuremath{\Ps}}\xspace}

\def\cquark    {{\ensuremath{\Pc}}\xspace}
\def\cquarkbar {{\ensuremath{\overline \cquark}}\xspace}

\def\pion   {{\ensuremath{\Ppi}}\xspace}

\def\pip    {{\ensuremath{\pion^+}}\xspace}

\def\kaon    {{\ensuremath{\PK}}\xspace}

\def\KorKbar {\kern \thebaroffset\optbar{\kern -\thebaroffset \PK}{}\xspace}

\def\Kp      {{\ensuremath{\kaon^+}}\xspace}
\def\Km      {{\ensuremath{\kaon^-}}\xspace}

\def\Dbar    {{\ensuremath{\offsetoverline{\PD}}}\xspace}
\def\D       {{\ensuremath{\PD}}\xspace}
\def\Db      {{\ensuremath{\Dbar}}\xspace}
\def\DorDbar {\kern \thebaroffset\optbar{\kern -\thebaroffset \PD}\xspace}
\def\Dz      {{\ensuremath{\D^0}}\xspace}
\def\Dzb     {{\ensuremath{\Dbar{}^0}}\xspace}
\def\Dp      {{\ensuremath{\D^+}}\xspace}
\def\Dm      {{\ensuremath{\D^-}}\xspace}
\def\Dpm     {{\ensuremath{\D^\pm}}\xspace}

\def\DpDm    {\ensuremath{\Dp {\kern -0.16em \Dm}}\xspace}

\def\Ds      {{\ensuremath{\D^+_\squark}}\xspace}
\def\Dsp     {{\ensuremath{\D^+_\squark}}\xspace}
\def\Dsm     {{\ensuremath{\D^-_\squark}}\xspace}
\def\Dspm    {{\ensuremath{\D^{\pm}_\squark}}\xspace}

\def\B       {{\ensuremath{\PB}}\xspace}

\def\BorBbar {\kern \thebaroffset\optbar{\kern -\thebaroffset \PB}\xspace}

\def\Bd      {{\ensuremath{\B^0}}\xspace}

\def\BdorBdbar {\kern \thebaroffset\optbar{\kern -\thebaroffset \Bd}\xspace}

\def\Bs      {{\ensuremath{\B^0_\squark}}\xspace}

\def\BsorBsbar {\kern \thebaroffset\optbar{\kern -\thebaroffset \Bs}\xspace}

\def\jpsi     {{\ensuremath{{\PJ\mskip -3mu/\mskip -2mu\Ppsi}}}\xspace}

\def\Y#1S{\ensuremath{\PUpsilon{(#1S)}}\xspace}

\def\LorLbar     {\kern \thebaroffset\optbar{\kern -\thebaroffset \PLambda}\xspace}

\def\BF         {{\ensuremath{\mathcal{B}}}\xspace}

\newcommand{\decay}[2]{\ensuremath{#1\!\to #2}\xspace} 

\def\to                 {\ensuremath{\rightarrow}\xspace}

\def\AT#1     {\ensuremath{A_{\mathrm{T}}^{#1}}\xspace}           

\def\C#1      {\ensuremath{\mathcal{C}_{#1}}\xspace}                       
\def\Cp#1     {\ensuremath{\mathcal{C}_{#1}^{'}}\xspace}                    
\def\Ceff#1   {\ensuremath{\mathcal{C}_{#1}^{\mathrm{(eff)}}}\xspace}        
\def\Cpeff#1  {\ensuremath{\mathcal{C}_{#1}^{'\mathrm{(eff)}}}\xspace}       
\def\Ope#1    {\ensuremath{\mathcal{O}_{#1}}\xspace}                       
\def\Opep#1   {\ensuremath{\mathcal{O}_{#1}^{'}}\xspace}                    

\newcommand{\aunit}[1]{\ensuremath{\text{\,#1}}}       

\newcommand{\tev}{\aunit{Te\kern -0.1em V}\xspace}
\newcommand{\gev}{\aunit{Ge\kern -0.1em V}\xspace}
\newcommand{\mev}{\aunit{Me\kern -0.1em V}\xspace}
\newcommand{\kev}{\aunit{ke\kern -0.1em V}\xspace}
\newcommand{\ev}{\aunit{e\kern -0.1em V}\xspace}
 
\newcommand{\mevc}{\ensuremath{\aunit{Me\kern -0.1em V\!/}c}\xspace}
\newcommand{\gevc}{\ensuremath{\aunit{Ge\kern -0.1em V\!/}c}\xspace}
\newcommand{\mevcc}{\ensuremath{\aunit{Me\kern -0.1em V\!/}c^2}\xspace}
\newcommand{\gevcc}{\ensuremath{\aunit{Ge\kern -0.1em V\!/}c^2}\xspace}

\def\barn{\aunit{b}\xspace}
\def\mbarn{\aunit{mb}\xspace}

\def\nb {\aunit{nb}\xspace}
\def\invnb {\ensuremath{\nb^{-1}}\xspace}

\newcommand{\chisq}{\ensuremath{\chi^2}\xspace}

\def\gsim{{~\raise.15em\hbox{$>$}\kern-.85em
          \lower.35em\hbox{$\sim$}~}\xspace}
\def\lsim{{~\raise.15em\hbox{$<$}\kern-.85em
          \lower.35em\hbox{$\sim$}~}\xspace}

\def\sPlot{\mbox{\em sPlot}\xspace}

\def\sqs   {\ensuremath{\protect\sqrt{s}}\xspace}
\def\sqsnn {\ensuremath{\protect\sqrt{s_{\scriptscriptstyle\text{NN}}}}\xspace}
\def\pt         {\ensuremath{p_{\mathrm{T}}}\xspace}

\newcommand{\lum} {\ensuremath{\mathcal{L}}\xspace}

\def\evtgen     {\mbox{\textsc{EvtGen}}\xspace}

\def\geant      {\mbox{\textsc{Geant4}}\xspace}

\def\tell1  {TELL1\xspace}
\def\ukl1   {UKL1\xspace}

\newcommand{\ie}{\mbox{\itshape i.e.}\xspace}

\def\DD {\ensuremath{\mathrm{LS}}\xspace}
\def\DDb {\ensuremath{\mathrm{OS}}\xspace}
\def\LS {\ensuremath{\mathrm{LS}}\xspace}
\def\OS {\ensuremath{\mathrm{OS}}\xspace}

\def\DPS {\ensuremath{\mathrm{DPS}}\xspace}

\newcommand\protonLead {$p$-Pb\xspace}
\def\pPb {\ensuremath{p\mathrm{Pb}}\xspace}

\def\Pbp {\ensuremath{\mathrm{Pb}p}\xspace}
\def\sigmaEff {\ensuremath{\sigma_\mathrm{eff}}\xspace}
\def\sigmaEffpp {\ensuremath{\sigma_{\mathrm{eff},\,pp}}\xspace}
\def\sigmaEffpPb {\ensuremath{\sigma_{\mathrm{eff},\,p\mathrm{Pb}}}\xspace}
\def\decayDzToKpi {\decay{\Dz}{\Km\pip}}
\def\decayDpToKpipi {\decay{\Dp}{\Km\pip\pip}}
\def\decayDsToKKpi {\decay{\Ds}{\Km\Kp\pip}}

\def\decayJpsiToMuMu {\decay{\jpsi}{\mumu}}

\newcommand\CfB {charm-from-$b$\xspace}
\def\RFB      {\ensuremath{R_{\mathrm{FB}}}\xspace}

\def\RpPb      {\ensuremath{R}\xspace}

\def\deltaPhi      {\ensuremath{\Delta \phi}\xspace}
\def\mDD {\ensuremath{m_{DD}}\xspace}
\def\pp {\ensuremath{pp}\xspace}

\newcommand{\pair}[2]{\ensuremath{#1#2}\xspace} 

\def\floatwidth{\columnwidth}

\usepackage{cite} 
\usepackage{mciteplus}

\usepackage{longtable} 

\begin{document}

\renewcommand{\thefootnote}{\fnsymbol{footnote}}
\setcounter{footnote}{1}

\onecolumn

\begin{titlepage}
\pagenumbering{roman}

\vspace*{-1.5cm}
\centerline{\large EUROPEAN ORGANIZATION FOR NUCLEAR RESEARCH (CERN)}
\vspace*{1.5cm}
\noindent
\begin{tabular*}{\linewidth}{lc@{\extracolsep{\fill}}r@{\extracolsep{0pt}}}
\ifthenelse{\boolean{pdflatex}}
{\vspace*{-1.5cm}\mbox{\!\!\!\includegraphics[width=.14\textwidth]{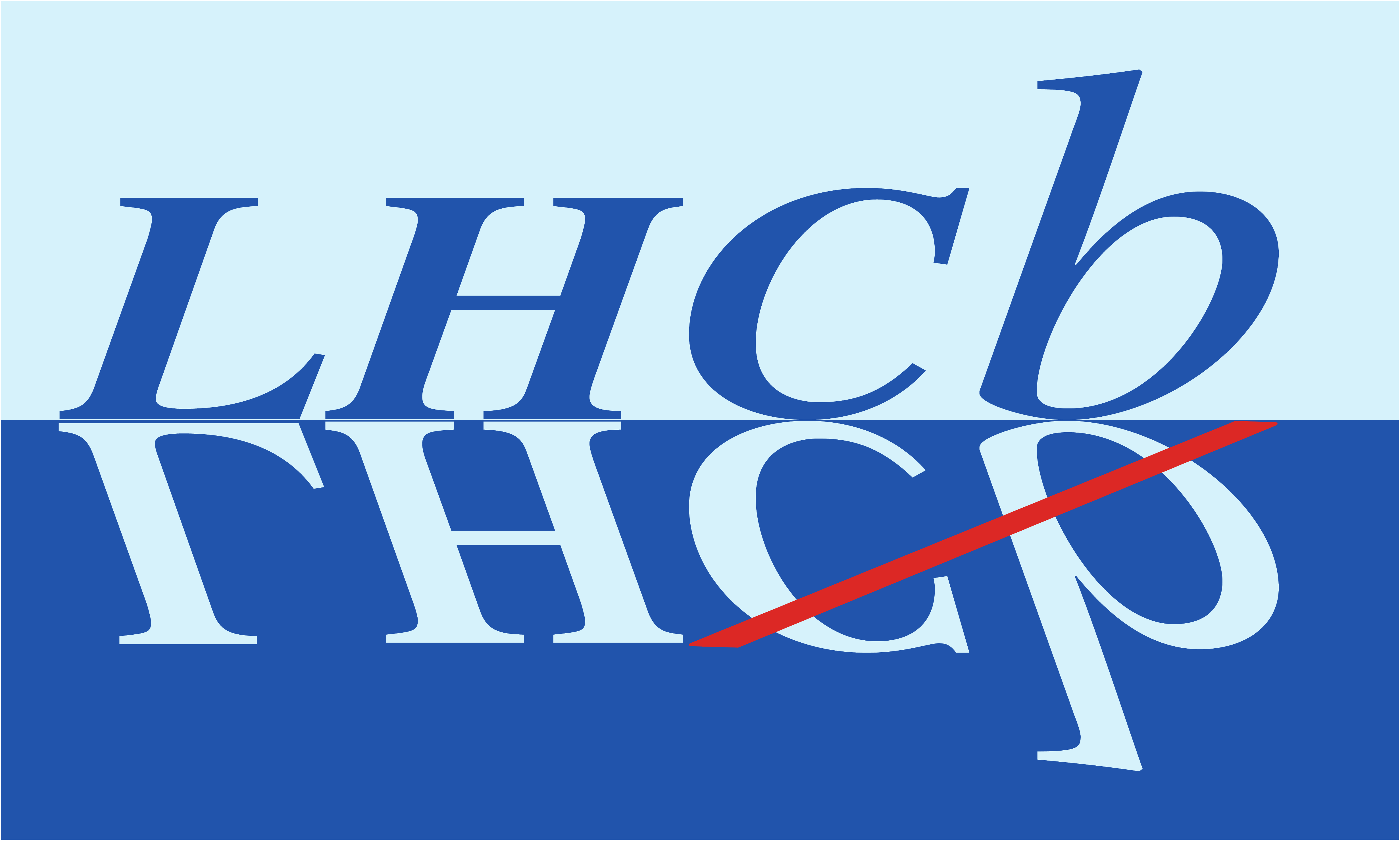}} & &}%
{\vspace*{-1.2cm}\mbox{\!\!\!\includegraphics[width=.12\textwidth]{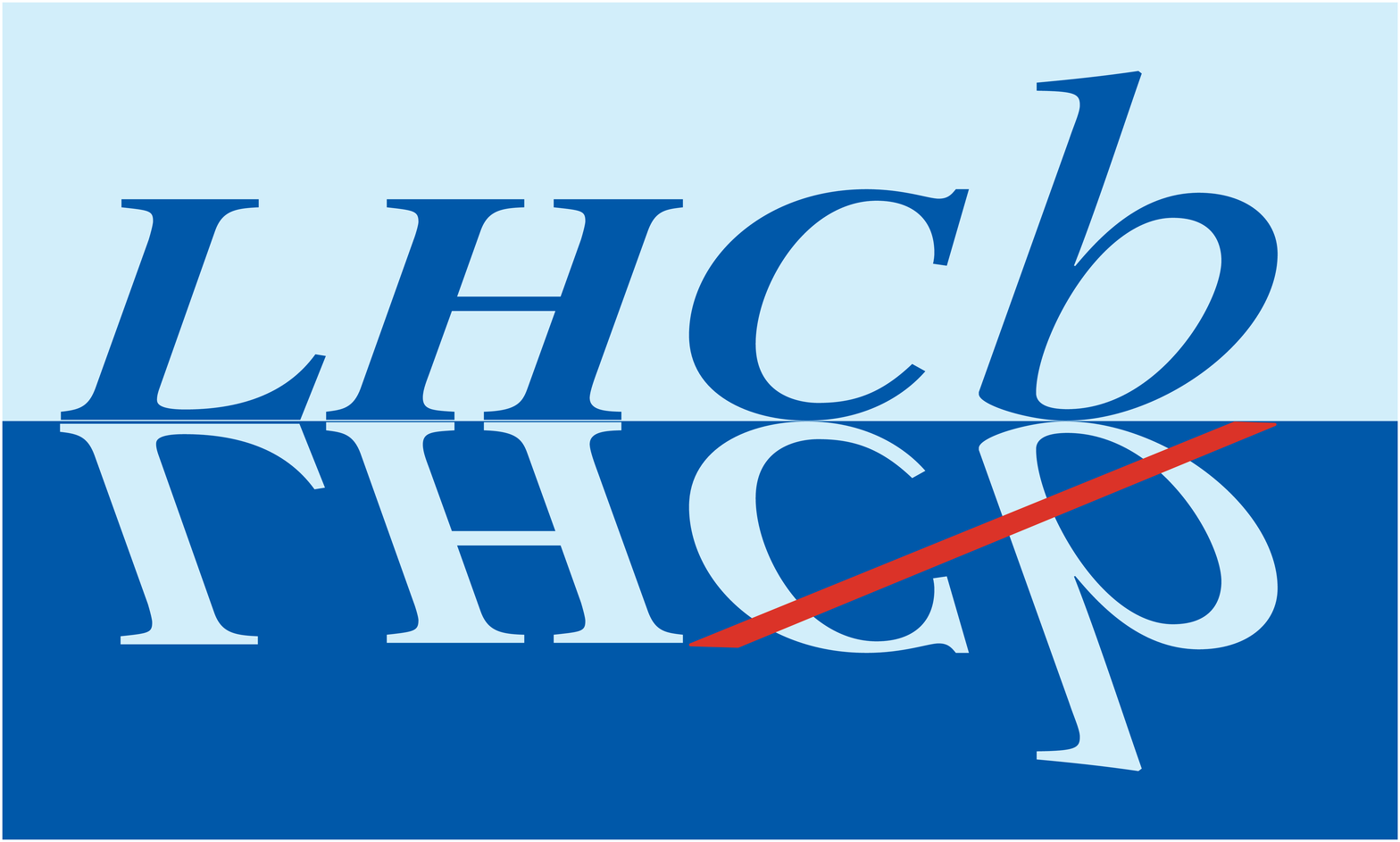}} & &}%
\\
 & & CERN-EP-2020-119 \\  
 & & LHCb-PAPER-2020-010 \\  
 & & \today \\ 
 & & \\
\end{tabular*}

\vspace*{4.0cm}

{\normalfont\bfseries\boldmath\huge
\begin{center}
  \papertitle 
\end{center}
}

\vspace*{2.0cm}

\begin{center}
\paperauthors\footnote{Authors are listed at the end of this paper.}
\end{center}

\vspace{\fill}

\begin{abstract}
  \noindent
A study of prompt charm-hadron pair production in proton-lead collisions at $\sqsnn=$~8.16~TeV is performed using data 
corresponding to an integrated luminosity of about $30\invnb$, collected with the LHCb experiment. 
Production cross-sections for different pairs of charm hadrons are measured and kinematic correlations between the two charm hadrons are investigated. 
This is the first measurement of associated production of two charm hadrons in proton-lead  collisions. 
The results confirm the predicted enhancement of double parton scattering production in proton-lead collisions compared
    to the single parton scattering production.
\end{abstract}

\vspace*{2.0cm}

\begin{center}
  Published in Phys. Rev. Lett. 125 (2020) 212001
\end{center}

\vspace{\fill}

{\footnotesize 
\centerline{\copyright~\papercopyright. \href{\paperlicenceurl}{\paperlicence}.}}
\vspace*{2mm}

\end{titlepage}


\newpage
\setcounter{page}{2}
\mbox{~}


\renewcommand{\thefootnote}{\arabic{footnote}}
\setcounter{footnote}{0}

\cleardoublepage


\pagestyle{plain} 
\setcounter{page}{1}
\pagenumbering{arabic}


At high energy hadron colliders, particles are produced in fundamental collisions of internal partons in the 
beam projectiles. The underlying parton densities are described by parton distribution functions (PDFs).
A collision event can produce multiple heavy-flavor hadrons via a single parton scattering (SPS) or multiple parton scatterings.
The latter, generating on average a larger number of charged tracks, could explain the heavy-flavor production rate in high-multiplicity events~\cite{Abramowicz:2013iva,Alice:Abelev:2012rz,Alice:Acharya:2019hao,Bartalini:2017jkk}. 
In a simple model, assuming that the PDFs of two partons in the same projectile are independent,
the associated production cross-section of final-state particles $A$ and $B$ from two separate partonic interactions, \ie a double parton scattering
(DPS) process, is related to the inclusive production cross-section of $A$ and
$B$, $\sigma^A$ and $\sigma^B$, as~\cite{Abe:1997xk,Seymour:2013sya,Gaunt:2009re,Gaunt:2010pi,Kom:2011bd,Luszczak:2011zp,Baranov:2011ch,Bartalini:2011jp,Lansberg:2019fgm,Alvioli:2019kcy}, 
\begin{equation}
    \sigma_\DPS^{AB} = \frac{1}{1+\delta_{AB}}\frac{\sigma^A\sigma^B}{\sigmaEff}. \label{eq:pocket}
\end{equation}
Here,  $\delta_{AB}=1$ if $A$ and $B$ are identical and is zero  otherwise, and $\sigmaEff$ is the so-called  effective
cross-section. The  parameter $\sigmaEff$ is related to the collision geometry and is expected to be independent of the final state~\cite{Ryskin:2011kk,Treleani:2007gi,Calucci:1999yz}.
In proton-ion collisions, following the Glauber model~\cite{Miller:2007ri}, SPS production cross-section is expected to scale with the ion mass
number in the absence of nuclear matter effects. 
However, DPS production is enhanced compared to a mass number scaling due to collisions of partons from two different nucleons in the ion, 
and the enhancement factor is about three in proton-lead (\protonLead) collisions~\cite{Strikman:2001gz,dEnterria:2012jam,Luszczak:2011zp,Calucci:2013pza,Cazaroto:2016nmu,dEnterria:2017yhd,Helenius:2019uge,Blok:2019fgg}.

The production of two open charm hadrons, $D_1D_2$, and  \pair{\jpsi}{\D} meson pairs is of particular interest in the
study of SPS and DPS processes, as the cross-section is relatively large and the high charm quark mass permits
perturbative calculations even at low transverse momentum (\pt). 
In this Letter, $D$ and $D_{1,2}$ refer to either a $\Dz$, $\Dp$ or $\Dsp$ meson and the inclusion of charge conjugate states is implied. Both like-sign (\DD) and opposite-sign (\DDb) open charm hadron pairs are considered. 
In an \DD pair the two hadrons have the same charm-quark flavor, while in an \DDb pair they have opposite charm flavors.
Pairs of \DDb charm hadrons can be produced from a $\cquark\cquarkbar$ pair via SPS, thus the kinematics of the two  hadrons are
correlated, while DPS produces correlated and uncorrelated \DDb pairs. The correlation in SPS production may be modified
in heavy-ion data compared to proton-proton (\pp) collisions,
due to nuclear matter effects~\cite{Vogt:2018oje,Vogt:2019xmm,Marquet:2017xwy,Albacete:2018ruq,LHCb-CONF-2018-005,Shi:2015cva,Cao:2015cba,Zhu:2006er}.
The \DDb correlation is predicted to be sensitive to the properties of the hot medium formed in ultra-relativistic heavy
nucleus-nucleus
collisions~\cite{Lang:2013wya,He:2014tga,Fries:2003vb,Fries:2003kq,Blaizot:2015hya,Cho:2019syk,Thews:2000rj,Andronic:2003zv,Greco:2003vf,Abelev:2012rv,Yao:2018sgn}. 

The two hadrons in an \DD pair produced in a DPS process are expected to be uncorrelated.
Studies of \DD pair production and correlation in different environments help to test the universality of the parameter \sigmaEff  and gain insight into the underlying parton
correlations~\cite{Blok:2012jr}.
Since  DPS production involves two parton pairs, it is very
sensitive to the nuclear PDF (nPDF) in proton-ion collisions, including its possible  dependence on the position inside the nucleus~\cite{Shao:2020acd}.

Production of \DDb charm and beauty pairs has been studied 
in fully reconstructed decays~\cite{Link:2003uj,Aitala:1998kh,Frabetti:1993jr,Barlag:1993qd,LHCb-PAPER-2012-003} and using 
partially reconstructed decays~\cite{Aidala:2018dqb,
Albajar:1993be,Abe:1998ac,Acosta:2004nj,Aaltonen:2007zza,Abbott:1999se, LHCb-PAPER-2017-020,Khachatryan:2011wq}, and  
the hadron and anti-hadron are found to be correlated; in particular, the azimuthal angle, $\deltaPhi$, between the two hadron directions projected to the plane transverse to the beam line  favors values
close to $\deltaPhi=0$ or $\pi$. 
Production of \LS charm pairs, double quarkonium and multiple jets at the \tevatron and the \lhc revealed  evidence of DPS signals~\cite{LHCb-PAPER-2011-013, LHCb-PAPER-2016-057,LHCb-PAPER-2012-003, LHCb-PAPER-2015-046,Khachatryan:2016ydm,Khachatryan:2014iia,Abazov:2015fbl,Abe:1993rv,Aaboud:2016dea,Abe:1997bp}. 
The effective cross-section is measured to be in the range of 10 to $20\mbarn$ for most final states, however, a value as low as $5\mbarn$ is extracted using double
quarkonium production~\cite{Aaboud:2016fzt,Lansberg:2017chq,Lansberg:2014swa}. More measurements are required to resolve this puzzle.

This Letter presents the first measurement of charm pair production 
in proton-lead  collisions at a nucleon-nucleon center-of-mass energy of $\sqsnn = 8.16\tev$.
The data were collected with the \lhcb experiment at a low interaction rate in two distinct beam configurations. 
In the $\pPb$ configuration, particles produced in the direction of the proton beam are analysed, while in the $\Pbp$ configuration particles are analysed in the Pb beam direction. The \pPb (\Pbp)  data correspond to an integrated luminosity of $12.2\pm0.3\invnb$ ($18.6\pm0.5\invnb$).
The detector  coordinate system is
defined to have  the $z$-axis aligned with the proton beam direction. In the following, particle rapidities ($y$) are defined
in the nucleon-nucleon rest frame.

The \lhcb detector is a single-arm forward spectrometer described in
detail in \mbox{Refs.~\cite{LHCb-DP-2008-001,LHCb-DP-2014-002}}. The online event selection is performed by a
trigger, which consists of a hardware stage, based on information from the calorimeter and muon systems, 
followed by a software stage, which applies a full event reconstruction.  
Charm hadrons ($H_c\equiv\Dz,\Dp,\Dsp,\jpsi$) are reconstructed online via the decays  \decayDzToKpi, \decayDpToKpipi, \decayDsToKKpi and
$\decayJpsiToMuMu$. The data samples are selected by the hardware trigger based on the calorimeter activity for  $D$
candidates and based on the muon system for \jpsi candidates.
Candidate pairs are formed by  \pair{\Dz}{\Dz}, \pair{\Dz}{\Dzb} and
\pair{\Dp}{\Dpm} combinations (same species), and  \pair{\Dz}{\Dpm}, \pair{\Dz}{\Dspm}, \pair{\Dp}{\Dspm} and
$\pair{\jpsi}{\D^{0,+}}$ combinations (different species). 
Other charm pairs are not considered due to their limited yield in the data.
The tracks used to reconstruct the $\D$ mesons are required to be positively identified as kaons or pions and must be
 separated from every primary \protonLead collision vertex (PV). These tracks are also required to
have $\pt>250\mevc$ and at least one track must have $\pt>500\mevc$ ($\pt>1000\mevc$) for $\Dz$
($\Dp,\Dsp$) final states. The tracks are required to form a vertex of good quality that is separated from every PV. The reconstructed \D mesons 
are required to be consistent with originating from a PV,
which favors prompt  production over mesons  from beauty-hadron  decays (denoted as \CfB).  
The two muons used to reconstruct \jpsi candidates are required to  have $\pt>500\mevc$ and form a good-quality vertex. 

In the offline selection, kaons and pions are  required to have momentum $p>3\gevc$, and muons to have $p>6\gevc$, $\pt>750\mevc$ and be positively identified by using information from all subdetectors~\cite{LHCb-DP-2018-001,LHCb-DP-2013-001}. 
The $\Km\Kp$ invariant mass from the  $\decayDsToKKpi$ decay is required to be within $\pm20\mevcc$ of the known
$\phi(1020)$ mass~\cite{PDG2019}. 
A kinematic fit is performed on each  charm hadron and on the pair, constraining them to originate from a  PV. Requirements on the fit qualities 
strongly reduce \CfB contributions but  retain more than 99\% of prompt pairs.

Results are obtained in a charm-hadron kinematic region $\pt(H_c)<12\gevc$ and  $1.7<y(H_c)<3.7$
($-4.7<y(H_c)<-2.7$) for \pPb (\Pbp) data.
For \Dp and \Dsp mesons the requirement $\pt(H_c)>2\gevc$ is applied due to extremely small yields at lower \pt.
Total cross-sections of \pair{\Dz}{\Dz}, \pair{\Dz}{\Dzb} and \pair{\jpsi}{\Dz} pair production
are also evaluated
in the full LHCb rapidity acceptance,
$1.5<y(H_c)<4$ ($-5<y(H_c)<-2.5$) for \pPb (\Pbp) data, in order to compare with single charm production~\cite{LHCb-PAPER-2017-014,
LHCb-CONF-2019-004}.

The cross-section for a charm pair is calculated as
\mbox{$\sigma=N^\mathrm{corr}/(\lum\times\BF_1\times\BF_2)$}, where $\lum$ is the integrated luminosity, 
and $N^\mathrm{corr}$ is the signal yield after efficiency correction and the subtraction of \CfB background.
The branching fractions of the two charm-hadron decays, $\BF_{1,2}$, are taken from Ref.~\cite{PDG2019} for the
$\Dz$, $\Dp$, $\jpsi$ decays, and  $\BF(\Dsp\to(\Kp\Km)_\phi\pip)=(2.24\pm0.13)\%$  from Refs.~\cite{Alexander:2008aa,LHCb-PAPER-2016-042}.
The raw signal yield is determined from an
unbinned maximum likelihood fit to the distribution of the invariant masses, $m_1$ and $m_2$, of the two charm hadrons.  
The two-dimensional probability densities comprise four components: signal-signal, background-background,
signal-background and background-signal for the first-second charm hadron in a pair. The background is mainly from random
combinations of tracks.
The signal component for each charm hadron is described by the sum of a Gaussian  and a Crystal Ball function (CB)~\cite{Skwarnicki:1986xj} 
and the background component by an exponential function. 
The distribution for pairs of same-species hadrons is constructed to be independent of the ordering of $m_1$ and $m_2$.
As an example, the $(m_1,m_2)$ distribution for  \pair{\Dz}{\Dz} candidates and its projection on $m_1$ and $m_2$ are are shown in
Fig.~\ref{fig:mass_DD} for \pPb data,  with the fit projections overlaid. 
More distributions are shown in the Supplemental Material. 
The  raw signal yield is between  100 and 4000 for all hadron pairs considered.

\begin{figure}[!tpb]
\centering
  \includegraphics[width=0.33\floatwidth]{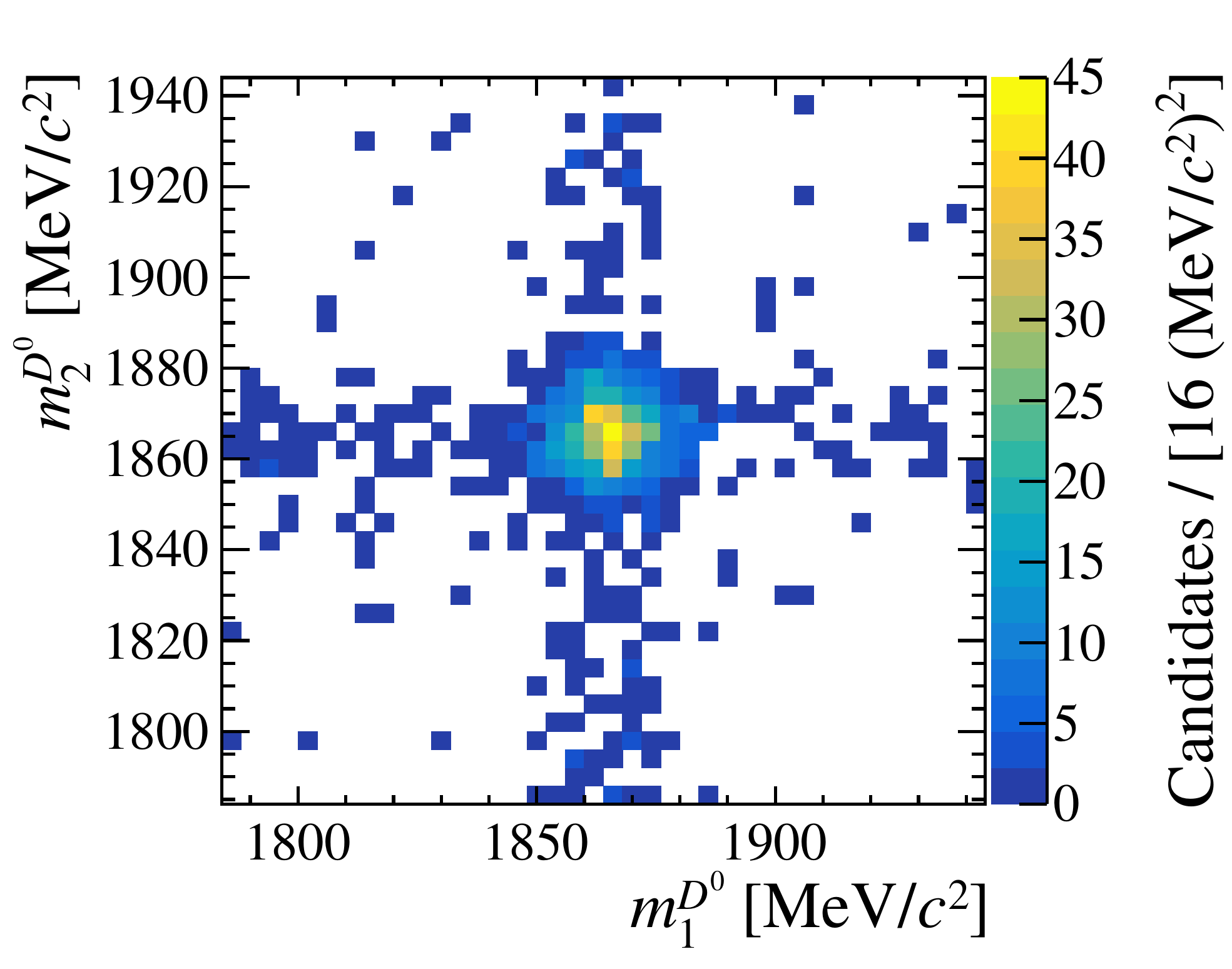}
  \includegraphics[width=0.66\floatwidth]{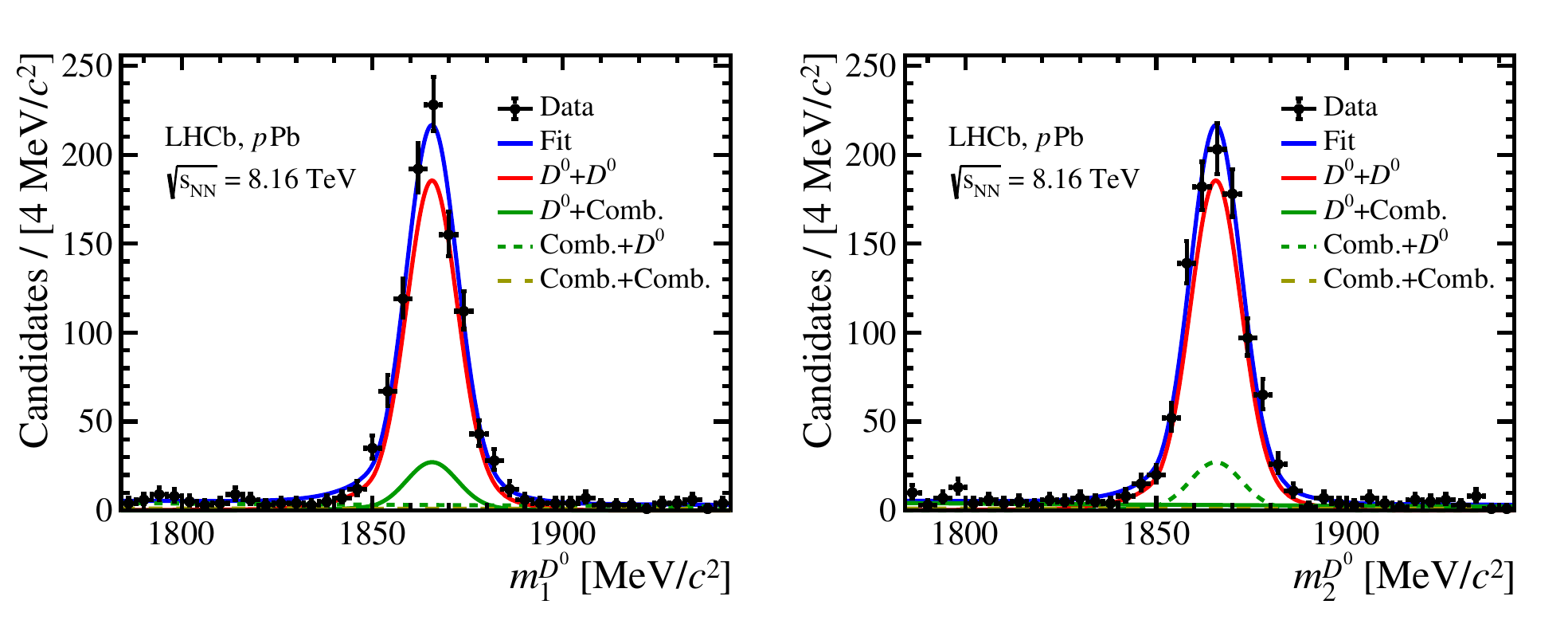}
  \caption{
      (Left) Two-dimensional invariant-mass distributions of $(m_1, m_2)$ for \pair{\Dz}{\Dz} pairs and the projections on (middle) $m_1$ and (right) $m_2$ with  the fit results superimposed. 
      Shown  in the projection plots are (points with bars) \pPb data, (solid blue) the total fit and its four  components.
    }
  \label{fig:mass_DD}
\end{figure} 

The total detection efficiency for each individual charm
hadron is evaluated from simulated signal decays, properly corrected using control samples of \protonLead collisions. These control samples
 are used to calibrate  track finding and particle identification (PID) efficiencies~\cite{LHCb-PAPER-2018-048}. 
In the simulation, minimum-bias \protonLead collisions are produced 
using the EPOS generator~\cite{Pierog:2013ria} 
according to beam configurations of the data. 
Charm hadrons are generated in \pp collisions at $\sqs=8.16\tev$
using \textsc{Pythia8}~\cite{Sjostrand:2007gs,Sjostrand:2006za} and are embedded into EPOS minimum-bias events.  
Particle decays are described by \evtgen~\cite{Lange:2001uf},
while the particle interaction with the detector, and its response,
are implemented using the \geant toolkit~\cite{Allison:2006ve, *Agostinelli:2002hh} as described in Ref.~\cite{LHCb-PROC-2011-006}.
The track finding efficiency in data and simulation is studied with a tag-and-probe method using $\jpsi\to\mumu$
decays~\cite{LHCb-DP-2013-002}. 
Similarly, the PID efficiency is measured
using large control samples of $\Dz\to\Km\pip$ and $\jpsi\to\mumu$ decays for $\Km,\pip$ and $\mun$ tracks, in bins of track momentum and pseudorapidity $(p,\eta)$. 
The average charged-track multiplicity in \OS data is similar to the one in the control samples, while for \LS data it is about 13\% higher, which is consistent with a larger contribution of multiple parton scattering in \LS data~\cite{Abramowicz:2013iva,Alice:Abelev:2012rz,Alice:Acharya:2019hao,Bartalini:2017jkk}.
The corresponding difference in detector occupancy results in different detection efficiencies in \LS and \OS data, which is evaluated in  control samples.
Efficiencies from control samples are combined with simulation to obtain
the efficiency for each charm hadron as a function of $\pt$ and $y$,  $\epsilon(\pt,y)$, which is used to determine the efficiency corrected signal yield
$\sum_i\frac{w^i}{\epsilon_1(\pt^i,y^i)\epsilon_2(\pt^i,y^i)}$. Here $w^i$ is the
signal \sPlot weight~\cite{Pivk:2004ty} used to  remove the contribution of background and is obtained from the fit to the invariant-mass distribution, and
$\epsilon_{1,2}(\pt^i,y^i)$ is the efficiency for the first and second hadron in the $i^{th}$
candidate pair in data.
The signal yield is then corrected for the \CfB contamination, 
which is estimated to be less than $1\%$ for open charm pairs and $(4\pm2)\%$ ($(3.0\pm1.5)\%$) for \pair{\jpsi}{\D} pairs in \pPb (\Pbp) data.

Several sources of systematic uncertainties are investigated.
The variation of the signal yield is studied with fits to the invariant-mass distribution using a different
signal or background model. 
A maximum relative variation of 2\% is obtained on the signal yield. 
The dominant systematic uncertainty arises from 
the limited control sample size to determine the track finding efficiency, 
which is on average about 5\% (10\%) per track in \pPb (\Pbp) data. 
An uncertainty of 2\% per hadron track is introduced to account for the loss of particles due to interactions with the detector
material.
Due to the small sample size and the choice of $(p,\eta)$ binning for each track, 
the PID efficiencies obtained from control samples 
introduce an uncertainty of less than 1\% on the total efficiency of each charm hadron.
Other contributions include the uncertainty on the total efficiency due to the size of the simulation sample, the uncertainty  
on the charm decay branching fractions, the uncertainty on the luminosity measurements and on the \CfB
fraction. These uncertainties are propagated to the cross-section measurements.

Total cross-sections are determined for all charm pairs. Results are detailed in the Supplemental
Material. For \LS open charm pairs, the measurements are in good agreement with theoretical calculations
including both SPS and DPS production~\cite{Helenius:2019uge}. The $\pair{\jpsi}{\Dz}$ cross-section is found to be
generally higher than  SPS production,  calculated using the weighted EPPS16
nPDF~\cite{shao:2015vga,Shao:2012iz,Kusina:2017gkz,Eskola:2016oht}.

Prompt single charm cross-sections in \pPb data were measured to be smaller than those of \Pbp data~\cite{LHCb-PAPER-2017-015,LHCb-CONF-2019-004}, which is explained by modifications of the nPDF. The same effect would result in even stronger suppression of DPS production in \pPb  compared to \Pbp data due to the participation of two pairs of partons. For charm pairs, the cross-section ratio between \pPb and \Pbp data,  the forward-backward ratio ($\RFB$), is determined for
$2.7<|y(H_c)|<3.7, \pt(H_c)>2\gevc$, to be 
$0.40\,\pm0.05\,\pm0.10$ ($0.61\,\pm0.04\,\pm0.12$) averaged over  \LS (\OS) open charm pairs, 
and is $0.26\,\pm0.06\,\pm0.04$ for \pair{\jpsi}{\D} pairs. Here and in the following, the first uncertainty is statistical and the second is systematic.
The results indicate reduced production in \pPb  compared to \Pbp data for both \LS and \OS pairs. 
The \RFB of \OS production is
compatible with that of prompt \Dz mesons~\cite{LHCb-PAPER-2017-015,LHCb-CONF-2019-004}, while that of \LS production
is smaller.  The ratio between the \RFB of \LS and \OS production, $ 0.66\,\pm0.09\,\pm0.03$, is in good agreement with 
the \RFB of \OS data and the \RFB of prompt \Dz production. The measurements favor the interpretation of \LS production via DPS.

The \LS over \OS cross-section ratio, $R^{D_1D_2}\equiv\sigma^{D_1D_2}/\sigma^{D_1\Db_2}$, is determined for all studied $\pair{\D_1}{\D_2}$
pairs under the $\pt(D)>2\gevc$ requirement, giving an average value of $0.308\pm0.015\pm0.010$ and $0.391\pm0.019\pm0.025$ for \pPb and
\Pbp data respectively.
The measurements agree with the calculations in Ref.~\cite{Helenius:2019uge} of $0.57^{+0.16}_{-0.41}$ (\pPb) and
$0.52^{+0.17}_{-0.38}$ (\Pbp), and 
are significantly larger than that in \pp~collisions where  $R^{\Dz\Dz}=0.109\pm0.008$~\cite{LHCb-PAPER-2012-003}, indicating
an enhancement of \LS pair production over \OS pairs in \protonLead collisions. The differential results as a function of $y(H_c)$ is shown in the Supplemental Material.

\begin{figure}[!tb]
    \begin{center}
        \includegraphics[width=0.48\floatwidth]{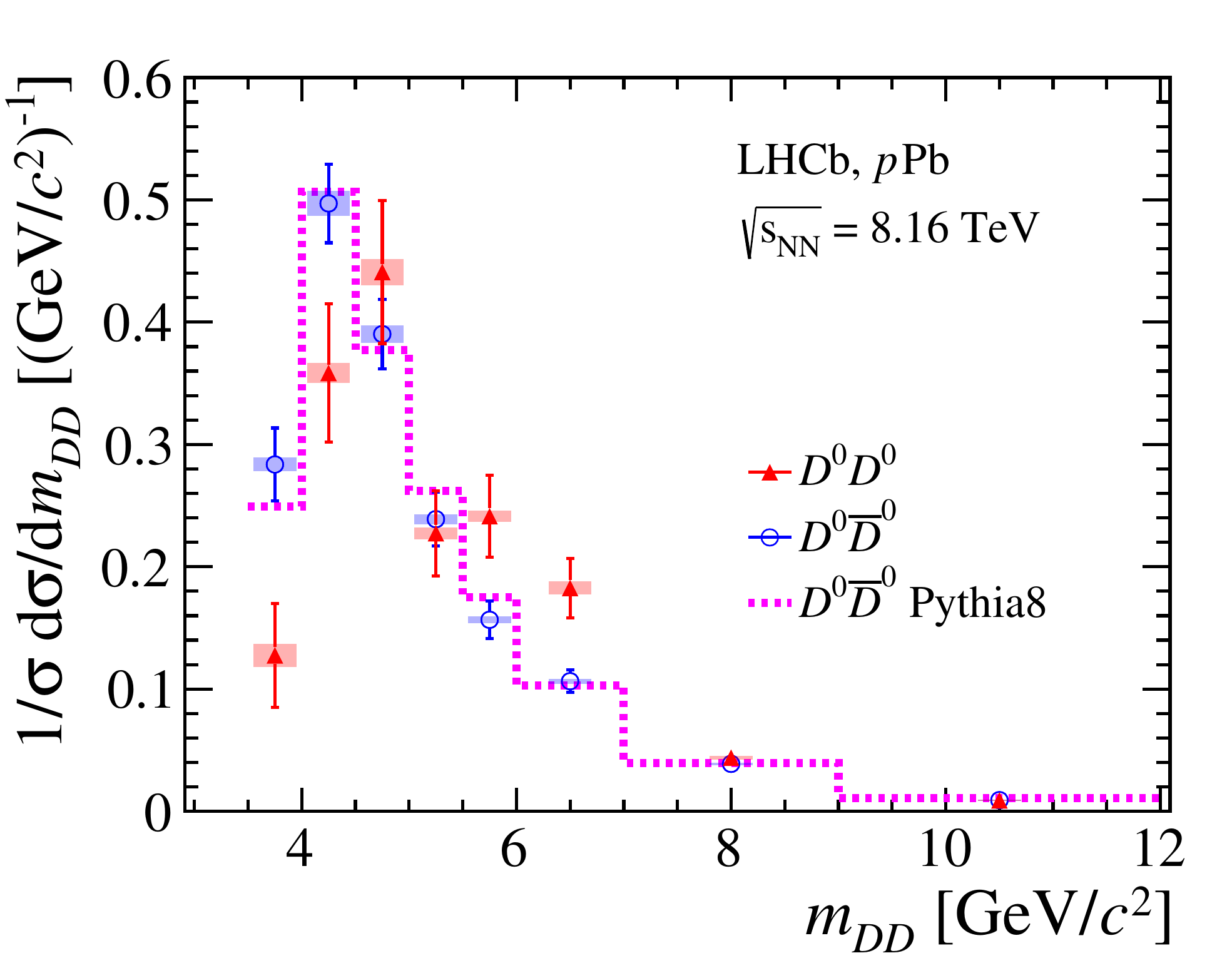}
        \includegraphics[width=0.48\floatwidth]{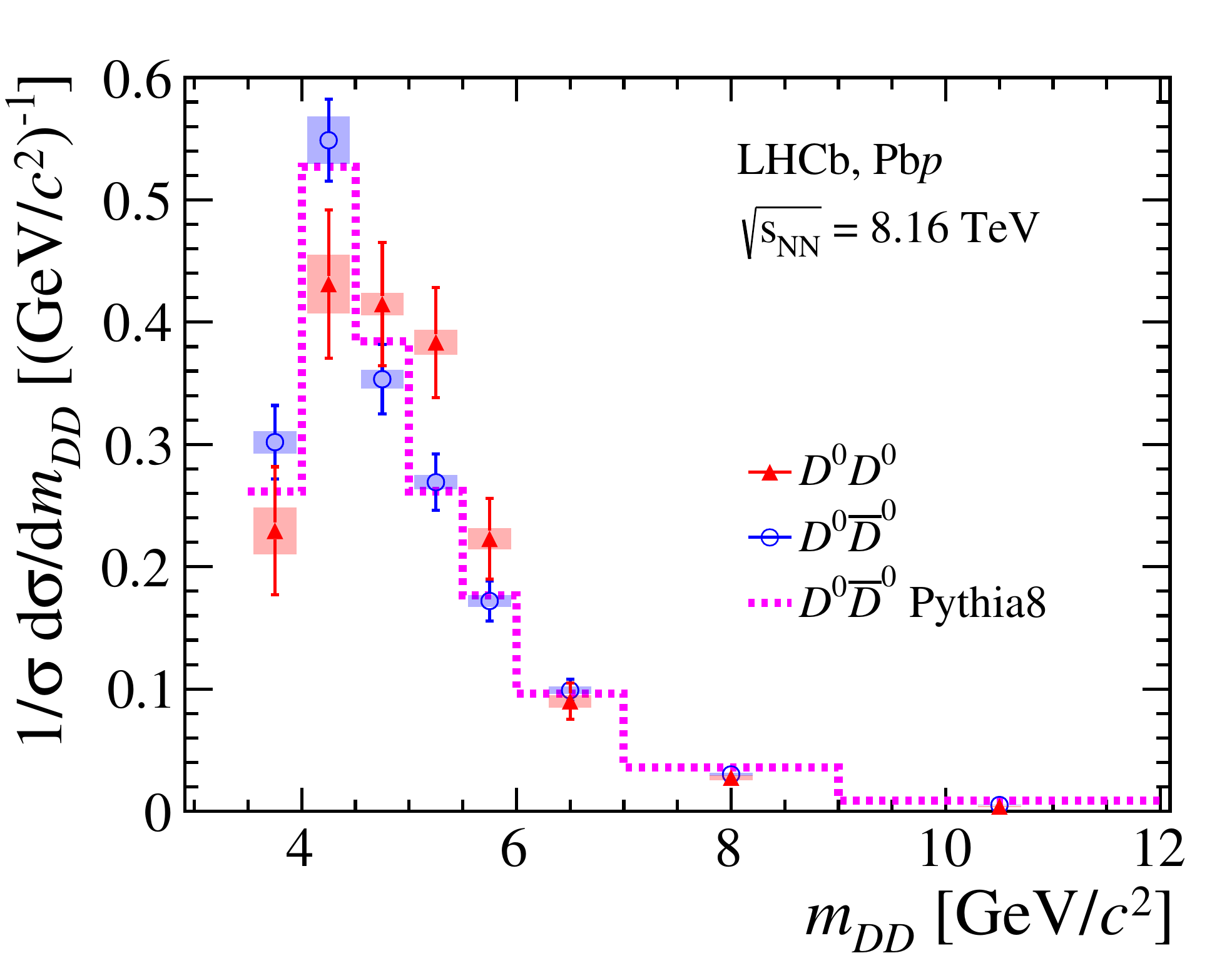}
    \end{center}
    \caption{
        Two-charm hadron invariant-mass  distribution of (red) $\pair{\Dz}{\Dz}$ and (blue) $\pair{\Dz}{\Dzb}$  pairs in (left) \pPb, (right) \Pbp data and
        (magenta dashed line) \textsc{Pythia8} simulation. 
    Vertical bars (filled box) are statistical (systematic) uncertainties.
    }
    \label{fig:XmD0D0}
\end{figure}

The correlations of kinematics between the two charm hadrons  in a pair are investigated from the distributions of the
two-charm invariant mass ($\mDD$) and their relative azimuthal angle $\deltaPhi$. 
The differential cross-section for each variable is normalized by the total cross-section, such that the largest systematic uncertainty, the one from the track finding efficiency,  almost completely cancels. As examples,
in Fig.~\ref{fig:XmD0D0}, the $\mDD$
distribution  is shown  for $\pair{\Dz}{\Dz}$  and $\pair{\Dz}{\Dzb}$
pairs without any requirement on $\pt(D)$. The difference between $\pair{\Dz}{\Dz}$ and $\pair{\Dz}{\Dzb}$ pairs
is determined to be
more than three (two) standard deviations in \pPb (\Pbp) data, studied using a \chisq test. 
For both $\pair{\Dz}{\Dz}$  and $\pair{\Dz}{\Dzb}$ pairs, the $\mDD$ distribution is compatible between \pPb and \Pbp data.
The  \pair{\Dz}{\Dzb} pair shows a similar $\mDD$ distribution to 
that of the  \textsc{Pythia8} simulation, in which the fraction of inclusive charm
production that contains more than one charm pair within the \lhcb acceptance is about 7\%.

\begin{figure}[!tpb]
    \begin{center}
        \includegraphics[width=0.48\floatwidth]{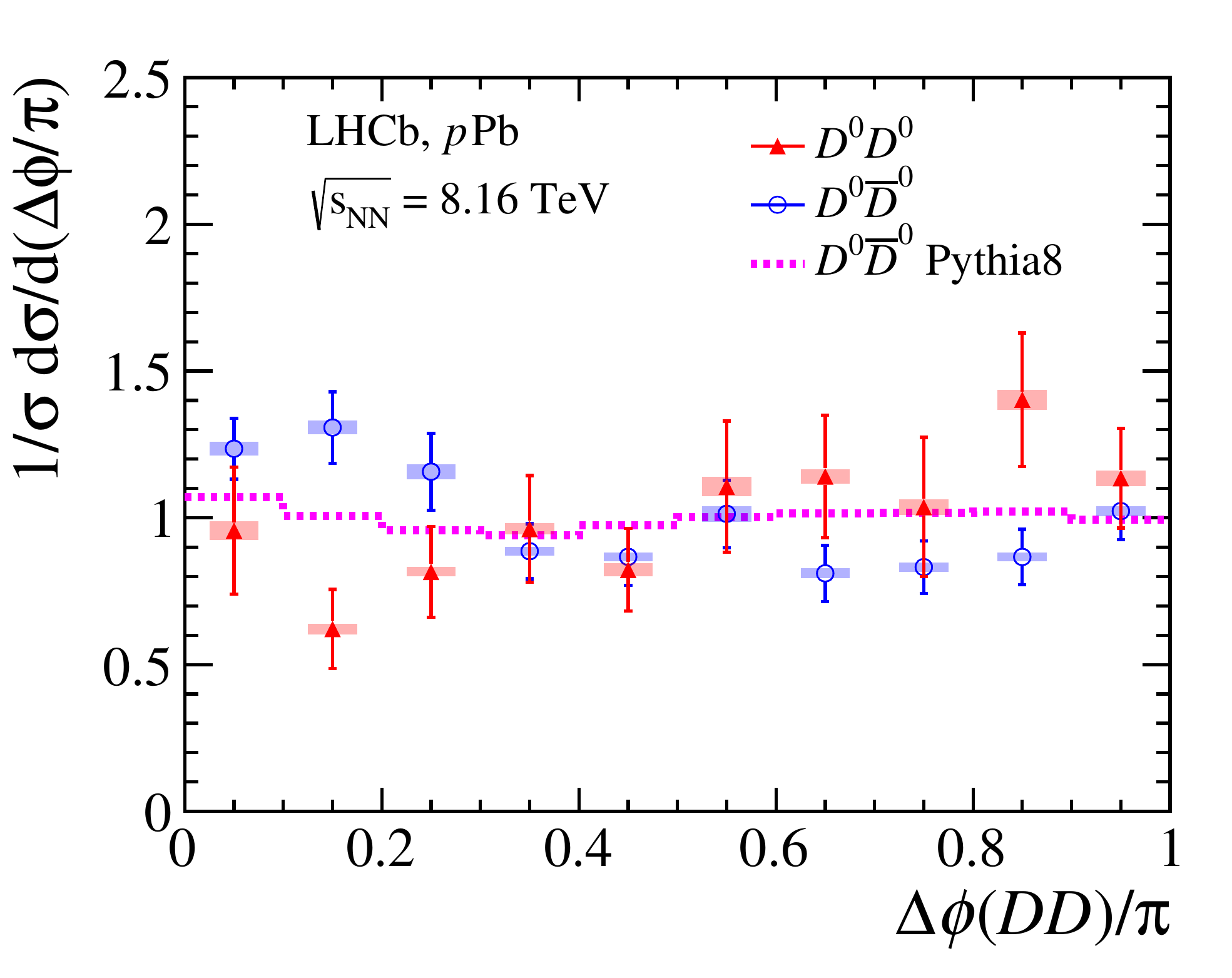}
        \includegraphics[width=0.48\floatwidth]{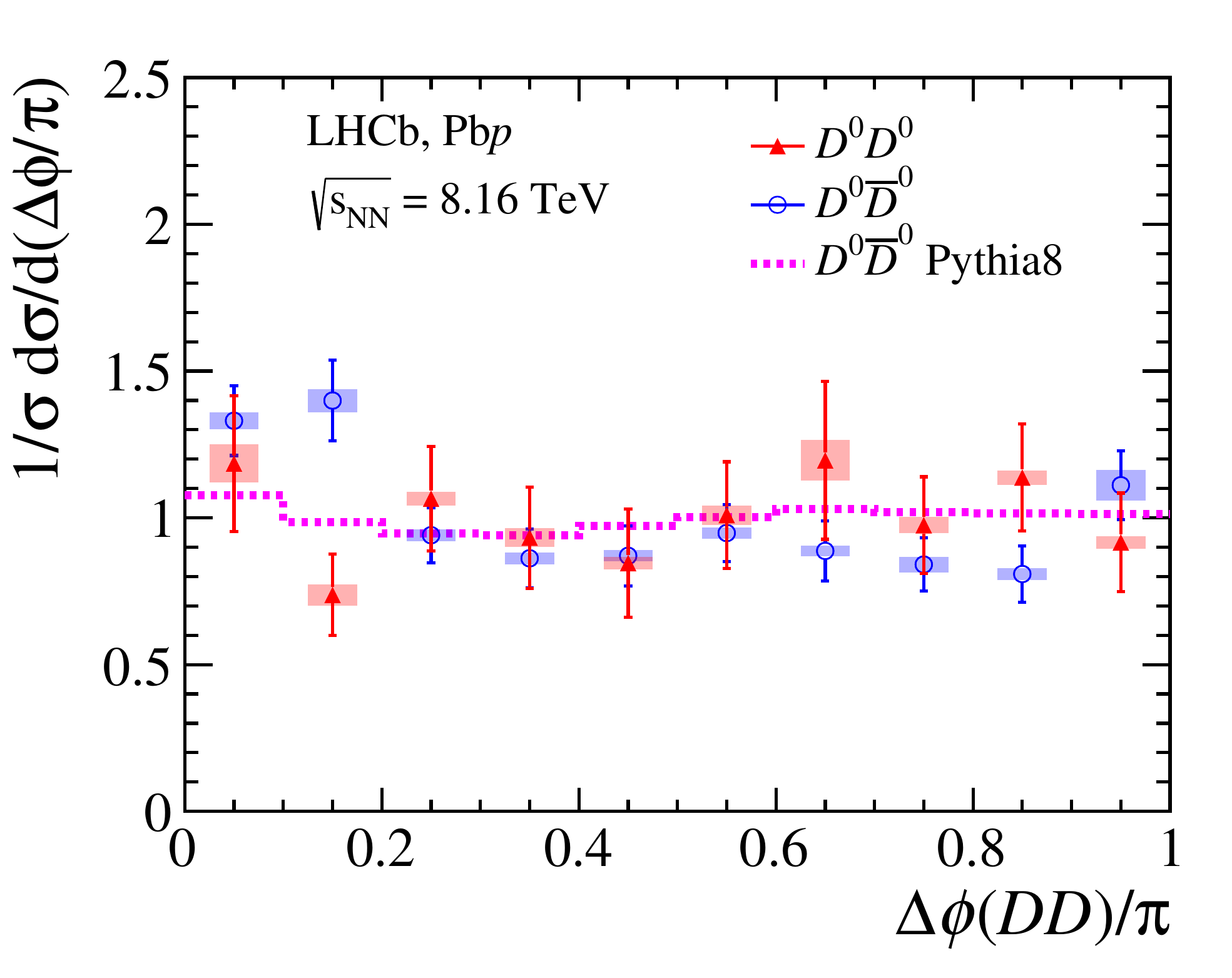}
        \includegraphics[width=0.48\floatwidth]{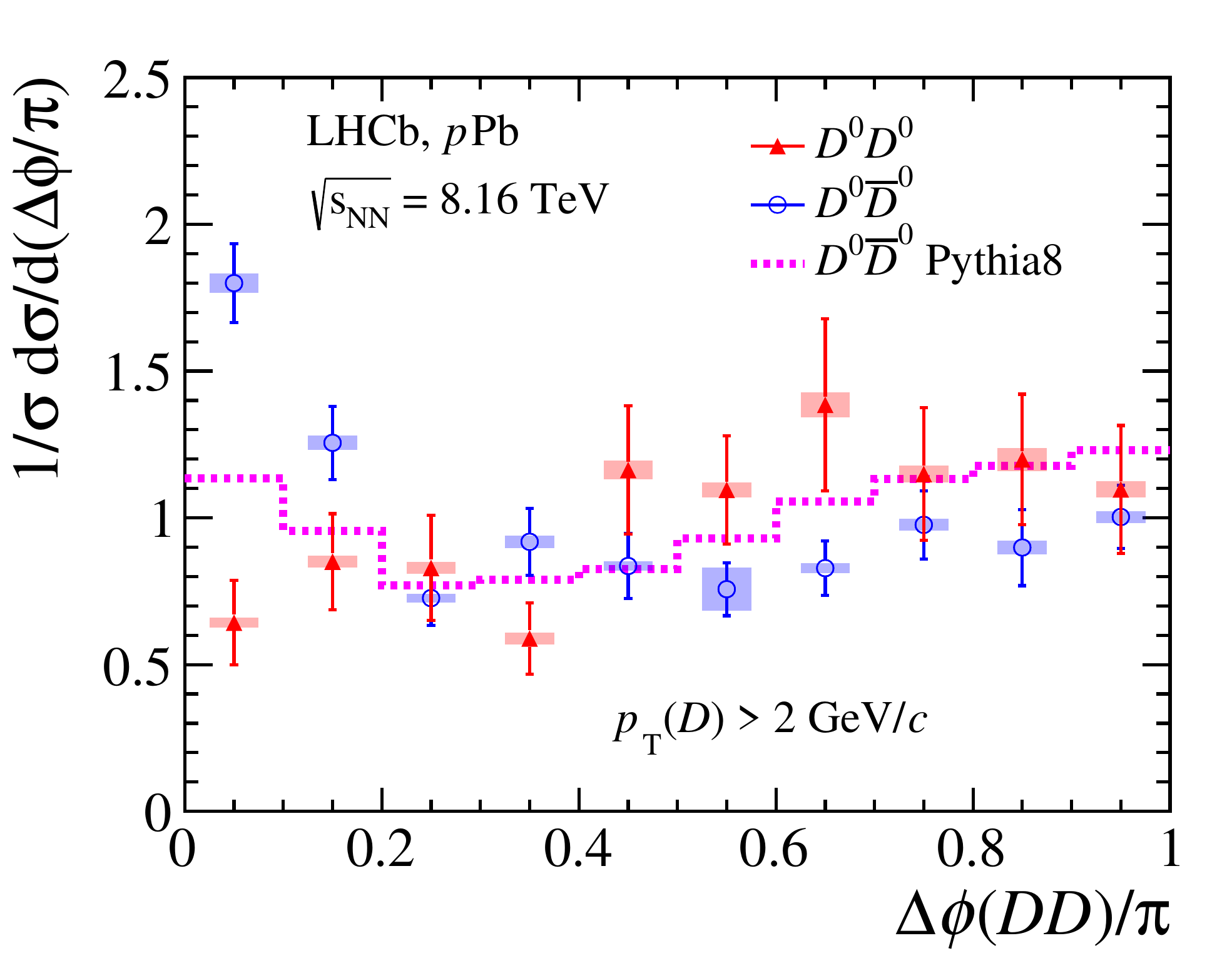}
        \includegraphics[width=0.48\floatwidth]{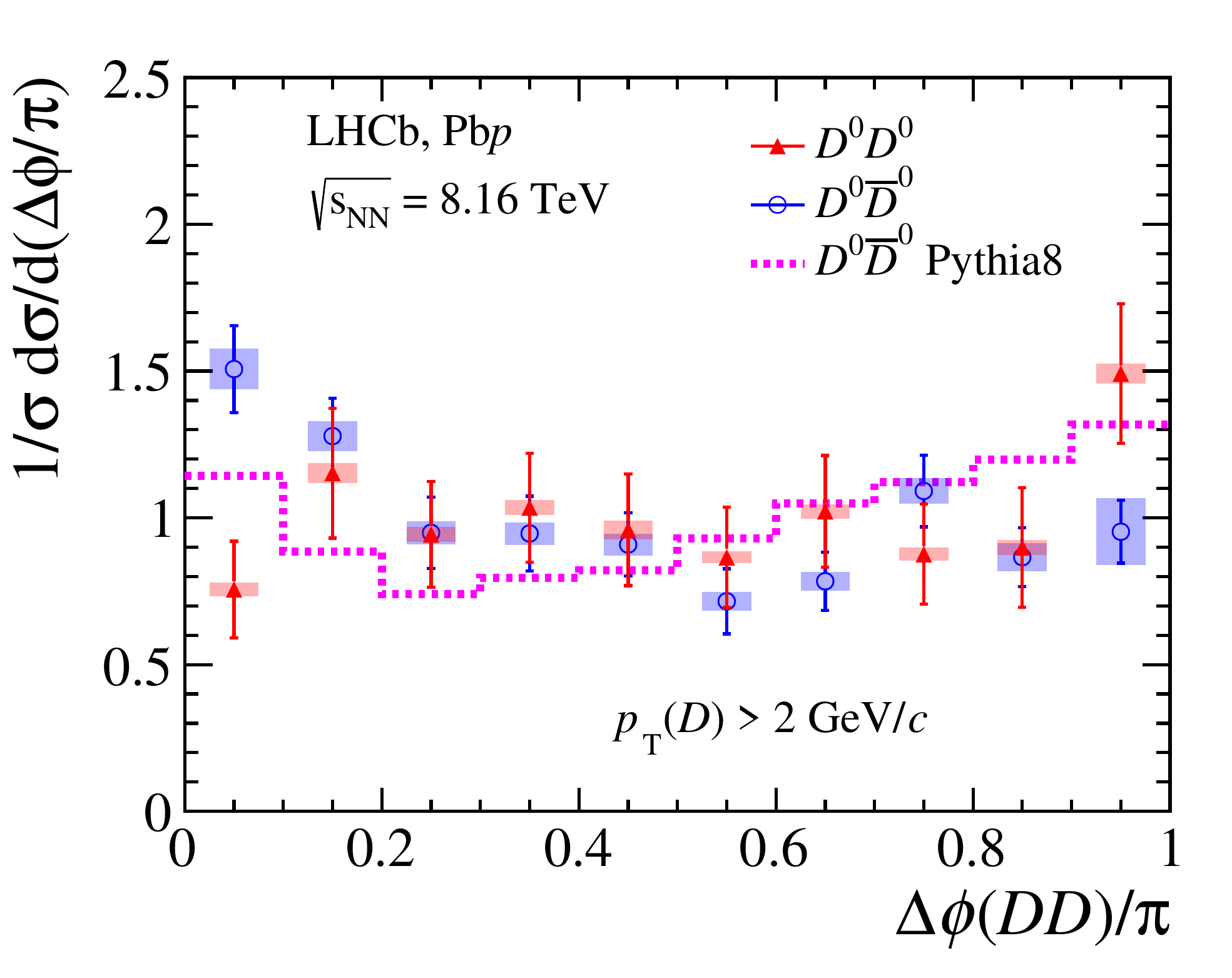}
    \end{center}
    \caption{
        The \deltaPhi distribution for (red) $\pair{\Dz}{\Dz}$ and (blue) $\pair{\Dz}{\Dzb}$ pairs in (left) \pPb, (right) \Pbp data and the 
        (magenta dashed line) \textsc{Pythia8} simulation, (bottom) with  and (top) without the \mbox{$\pt(\Dz)>2\gevc$}
        requirement. Vertical bars (filled box)
        are statistical (systematic) uncertainties.
    }
    \label{fig:dphiD0D0}
\end{figure}

The $\deltaPhi$ distribution
is shown in Fig.~\ref{fig:dphiD0D0} for $\pair{\Dz}{\Dz}$  and $\pair{\Dz}{\Dzb}$ pairs 
with and without the requirement $\pt(\Dz)>2\gevc$. 
Without this condition, the $\deltaPhi$ distribution is almost uniform for both  \LS and \OS pairs, similar to that in
\textsc{Pythia8} simulation. 
However, with the  $\pt(\Dz)>2\gevc$ requirement, the $\pair{\Dz}{\Dzb}$ pair favors values $\deltaPhi\sim0$,
while that of $\pair{\Dz}{\Dz}$ pairs is still compatible with being flat, and both show inconsistency with the \textsc{Pythia8}
simulation. 
In general, the behaviour that $\mDD$ distribution in $\pair{\Dz}{\Dz}$ pairs peaks at higher values compared to that of $\pair{\Dz}{\Dzb}$  pairs and 
the flat $\pair{\Dz}{\Dz}$ $\deltaPhi$
distribution are qualitatively consistent with a large DPS contribution in \LS pair production.
Distributions of the pair transverse momentum and the two-charm relative rapidity are found
 to be compatible  in \OS data, \LS data and the  \textsc{Pythia8} simulation.

\begin{table}[!bp]
    \centering
    \caption{
        The effective cross-section \sigmaEffpPb (in $\barn$) measured using \pair{\jpsi}{\Dz}  and \pair{\Dz}{\Dz}  pair
        production in \protonLead data and the extrapolated values from $pp$ data~\cite{LHCb-PAPER-2016-042}.
    }
\begin{tabular}{c|cc|c}
\hline
    Pairs& $-5<y(H_c)<-2.5$   & $1.5<y(H_c)<4$  & $pp$ extrapolation\\
\hline
    \pair{\Dz}{\Dz}& $0.99\pm0.09\pm0.09$ &   $1.41\pm0.11\pm0.10$  & $4.3\pm0.5$\\
\pair{\jpsi}{\Dz}& $0.64\pm0.10\pm0.06$ &   $0.92\pm0.22\pm0.06$ &$3.1\pm0.3$ \\
\hline
\end{tabular}
\label{tab:sigmaEff}
\end{table}

The effective cross-section $\sigmaEffpPb$ is calculated according to Eq.~\ref{eq:pocket} using the  \pair{\Dz}{\Dz} and
\pair{\jpsi}{\Dz}
cross-sections~\cite{Seymour:2013sya}, assuming solely DPS production,  
where the prompt \jpsi and \Dz production are evaluated from LHCb measurements~\cite{LHCb-PAPER-2017-014, LHCb-CONF-2019-004}. 
The results are  displayed in Table~\ref{tab:sigmaEff} with a typical value of order $1\barn$. 
Table~\ref{tab:sigmaEff}  (``$pp$ extrapolation'') also provides the $\sigmaEffpp$ result~\cite{LHCb-PAPER-2016-042} scaled by the Pb nucleus mass
number 208, which is valid under the assumption of SPS production and absence of nuclear modification.
The result confirms the expectation that DPS production in \protonLead collisions is enhanced by a factor of
three compared to SPS production, consistent with the expectation from the Glauber model. 
The $\sigmaEffpPb$ value measured using \pair{\jpsi}{\Dz}  production is
smaller than that observed in \pair{\Dz}{\Dz} production, as measured in $pp$ data~\cite{LHCb-PAPER-2016-042},
which may be due to SPS contamination~\cite{1798024} or more than expected $\pair{\jpsi}{\Dz}$ DPS production. The
\pPb data shows a higher $\sigmaEffpPb$ value compared to \Pbp data, which may suggest 
a complicated structure of the nPDF, as studied in Ref.~\cite{Shao:2020acd}.

The nuclear modification factor, $\RpPb\equiv\frac{\sigma_{\pPb}}{208\sigma_{pp}}$, is measured for \pair{\jpsi}{\Dz} and
\pair{\Dz}{\Dz} pairs with $\RpPb^{H_cH_c'} = \RpPb^{H_c}\times
\RpPb^{H_c'}\times \frac{208\sigmaEffpp}{\sigmaEffpPb}$, where $\sigma_{\pPb}$ and $\sigma_{pp}$ are the cross-sections of charm pairs in \protonLead and $pp$ collisions, respectively.
Assuming variations of $\RpPb$ and $\sigmaEffpp$ as a function of collision energy are small for \pt-integrated production, 
using measurements of $\sigmaEffpp$~\cite{LHCb-PAPER-2012-003}, $\RpPb^\jpsi$~\cite{LHCb-PAPER-2017-014}
and $\RpPb^\Dz$~\cite{LHCb-PAPER-2017-015},  $\RpPb^{\Dz\Dz} = 1.3\pm0.2$ ($4.2\pm0.8$) and $\RpPb^{\jpsi\Dz} = 1.5\pm0.5$ ($4.6\pm1.3$) for \pPb
(\Pbp) data are obtained, where the uncertainties are the total. The results are about a factor of three larger compared to that of  single \jpsi or \Dz hadron production~\cite{LHCb-PAPER-2017-014,LHCb-PAPER-2017-015}.

To summarise, the production of \LS and \OS open charm hadron pairs as well as \pair{\jpsi}{\D} pairs are studied
in \protonLead collisions at $\sqsnn=8.16\tev$ using fully reconstructed decays.
The cross-section ratio between \LS and \OS pairs is found to be a factor of three higher than that in $pp$ data.
The forward-backward ratio of \OS pairs is compatible with single charm production, while a smaller value is found for \LS pairs.
Distributions of the two-charm invariant mass and relative azimuthal angle
show a difference between \LS and \OS pairs, and the \LS pairs exhibit a flat relative azimuthal angle distribution 
independent of charm hadron \pt. 
The  effective cross-section and nuclear modification
factor for \pair{\jpsi}{\Dz} and \pair{\Dz}{\Dz} are in general compatible with the expected
enhancement factor of three  for DPS over SPS production ratio from $pp$ to \protonLead collisions.
This is the first direct observation of such an enhancement using \LS charm production
in \protonLead data. 
The \sigmaEffpPb result is different between \pPb and \Pbp data and between \pair{\jpsi}{\Dz} and
\pair{\Dz}{\Dz} pairs may suggest additional effects not considered yet, which deserve further investigation using future \lhcb data samples.

\section*{Acknowledgements}
%
%
\noindent We would like to thank  Hannu Paukkunen and Huasheng Shao for providing 
theoretical predictions and helpful discussions.  
We express our gratitude to our colleagues in the CERN
accelerator departments for the excellent performance of the LHC. We
thank the technical and administrative staff at the LHCb
institutes.
We acknowledge support from CERN and from the national agencies:
CAPES, CNPq, FAPERJ and FINEP (Brazil); 
MOST and NSFC (China); 
CNRS/IN2P3 (France); 
BMBF, DFG and MPG (Germany); 
INFN (Italy); 
NWO (Netherlands); 
MNiSW and NCN (Poland); 
MEN/IFA (Romania); 
MSHE (Russia); 
MinECo (Spain); 
SNSF and SER (Switzerland); 
NASU (Ukraine); 
STFC (United Kingdom); 
DOE NP and NSF (USA).
We acknowledge the computing resources that are provided by CERN, IN2P3
(France), KIT and DESY (Germany), INFN (Italy), SURF (Netherlands),
PIC (Spain), GridPP (United Kingdom), RRCKI and Yandex
LLC (Russia), CSCS (Switzerland), IFIN-HH (Romania), CBPF (Brazil),
PL-GRID (Poland) and OSC (USA).
We are indebted to the communities behind the multiple open-source
software packages on which we depend.
Individual groups or members have received support from
AvH Foundation (Germany);
EPLANET, Marie Sk\l{}odowska-Curie Actions and ERC (European Union);
A*MIDEX, ANR, Labex P2IO and OCEVU, and R\'{e}gion Auvergne-Rh\^{o}ne-Alpes (France);
Key Research Program of Frontier Sciences of CAS, CAS PIFI, and the Thousand Talents Program (China);
RFBR, RSF and Yandex LLC (Russia);
GVA, XuntaGal and GENCAT (Spain);
the Royal Society
and the Leverhulme Trust (United Kingdom).

\clearpage
\setcounter{figure}{3}
\setcounter{table}{1}
\setcounter{equation}{1}
\setcounter{page}{1}
{\noindent\LARGE Supplemental Material}
\label{sec:Supplemental}
\section*{Invariant mass distributions and fit projections}
\begin{figure}[!pb]
\centering
  \includegraphics[width=0.33\floatwidth]{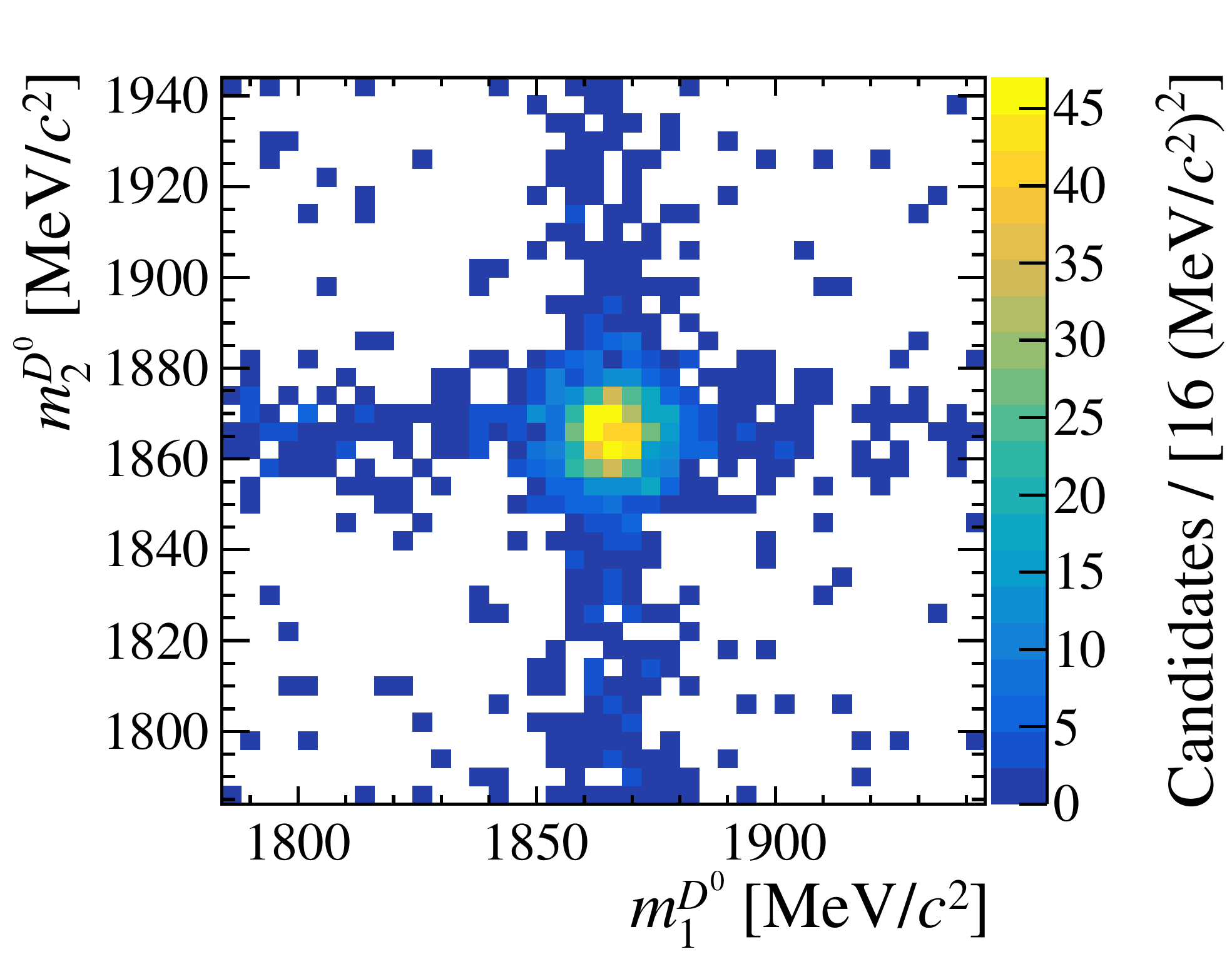}
  \includegraphics[width=0.66\floatwidth]{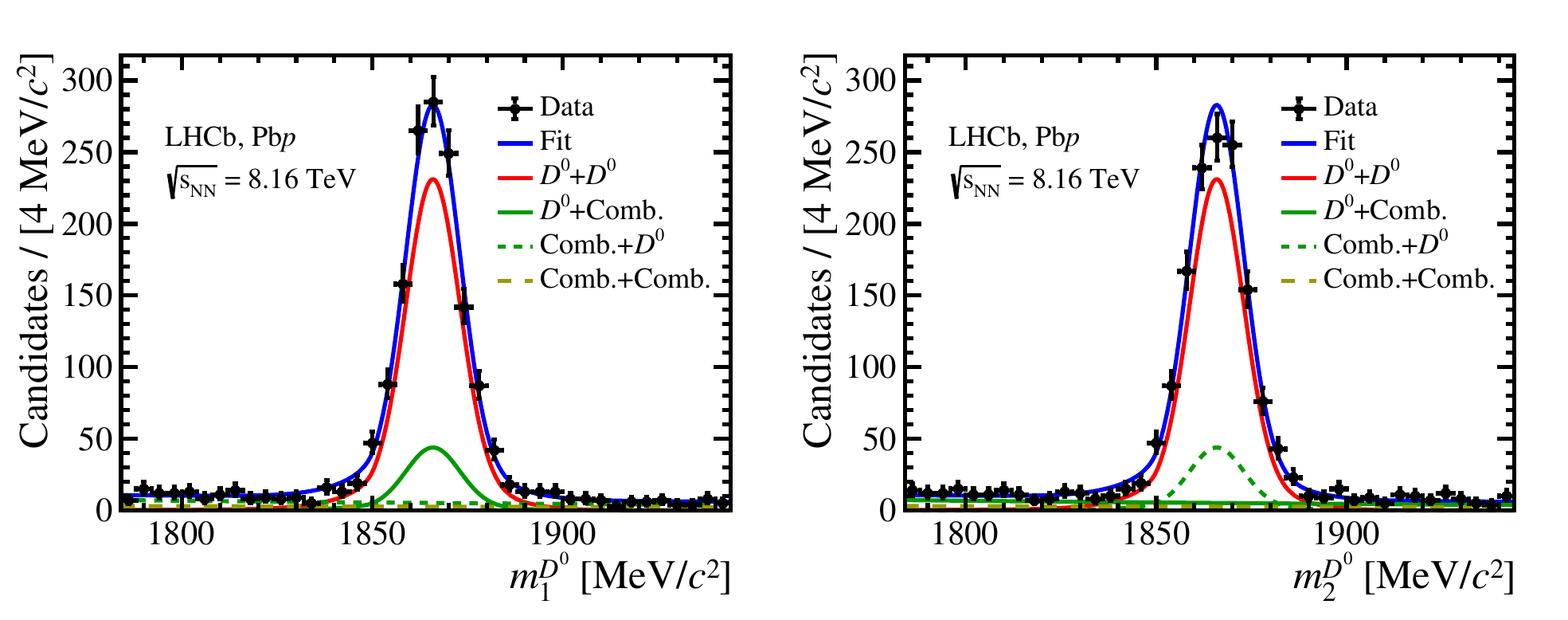}
  \caption{
  (Left) Two-dimensional invariant-mass distributions of $(m_1, m_2)$ for \pair{\Dz}{\Dz} pairs and the projections on (middle) $m_1$ and (right) $m_2$ with  the fit results superimposed. 
      Shown  in the projection plots are (points with error bars) \Pbp  data, (solid blue) the total fit and the four fit  components.
    }
  \label{fig:mass_DD_Pbp}
\end{figure}

\begin{figure}[!pbt]
\centering
\includegraphics[width=0.33\floatwidth]{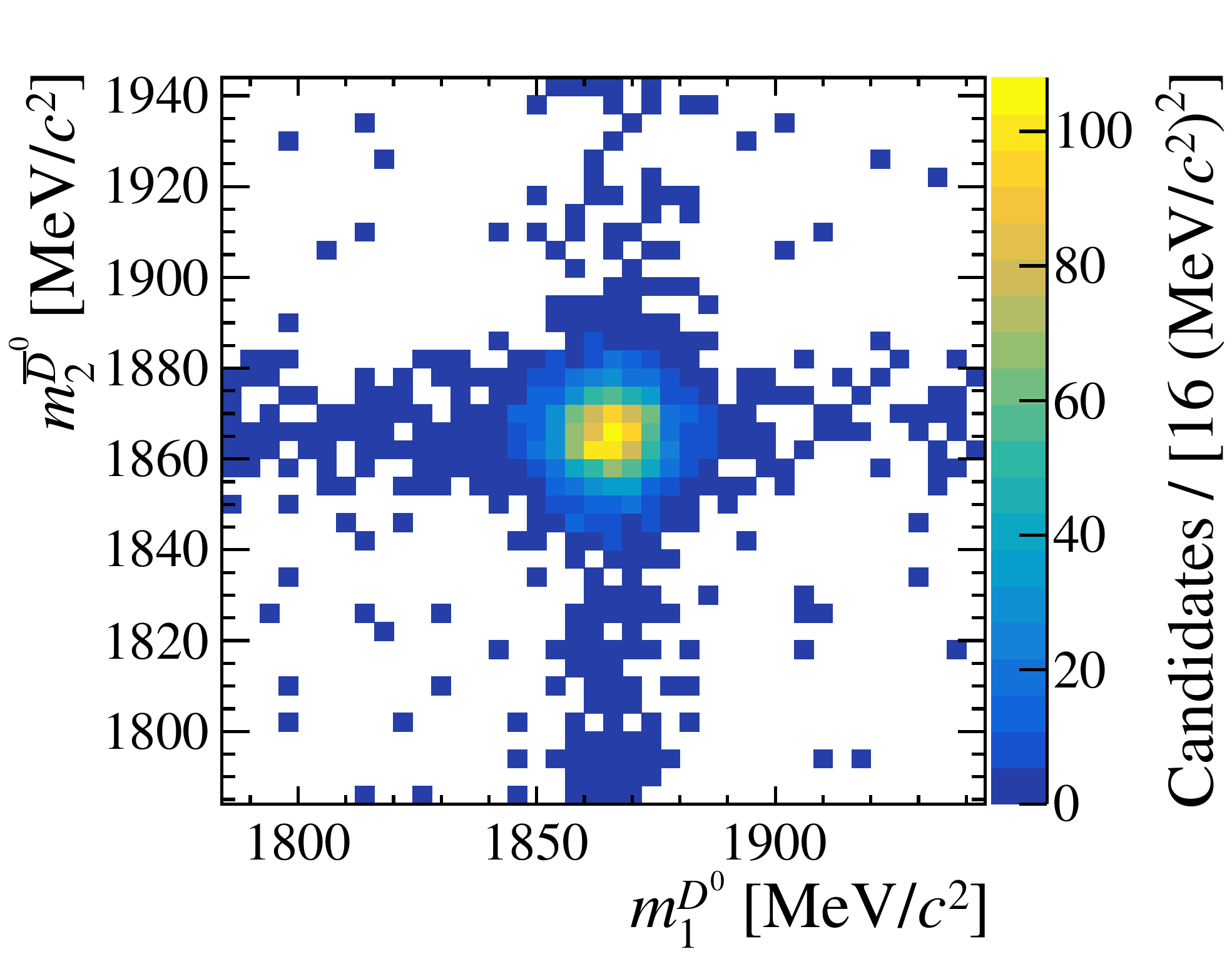}
  \includegraphics[width=0.66\floatwidth]{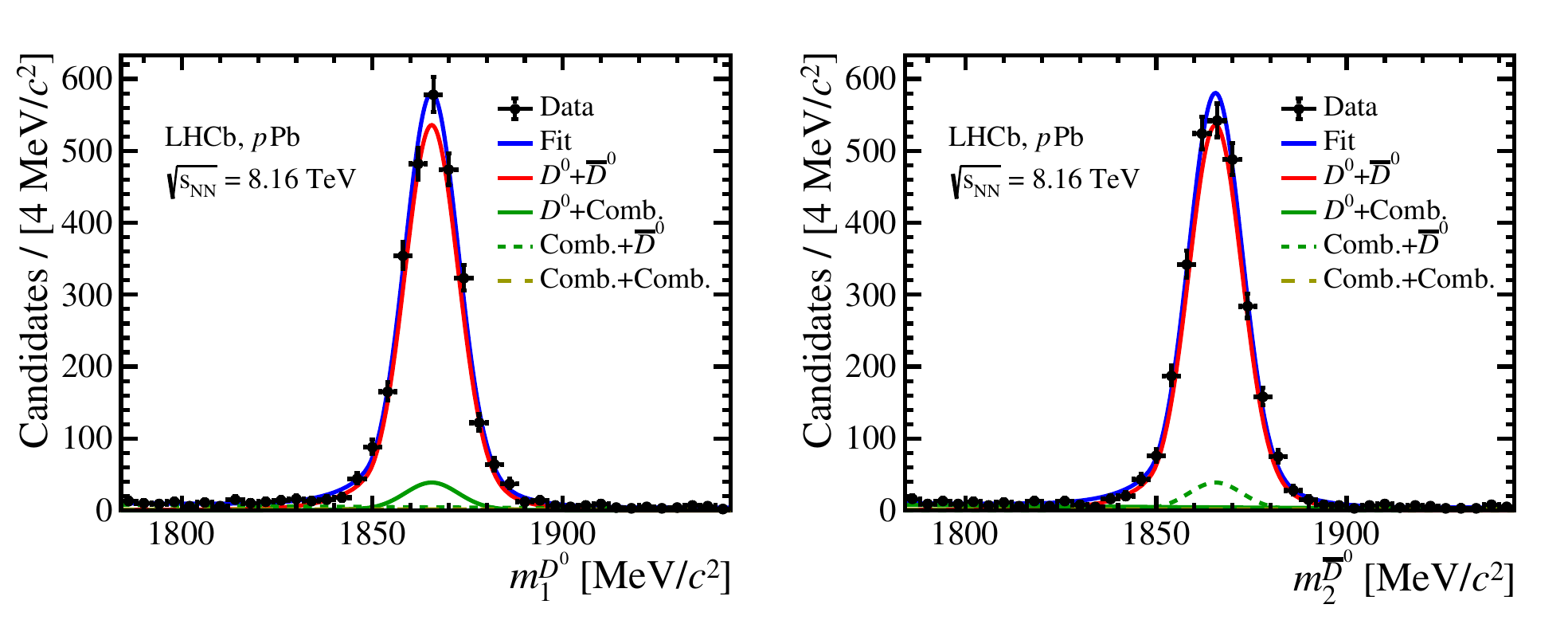}
   \includegraphics[width=0.33\floatwidth]{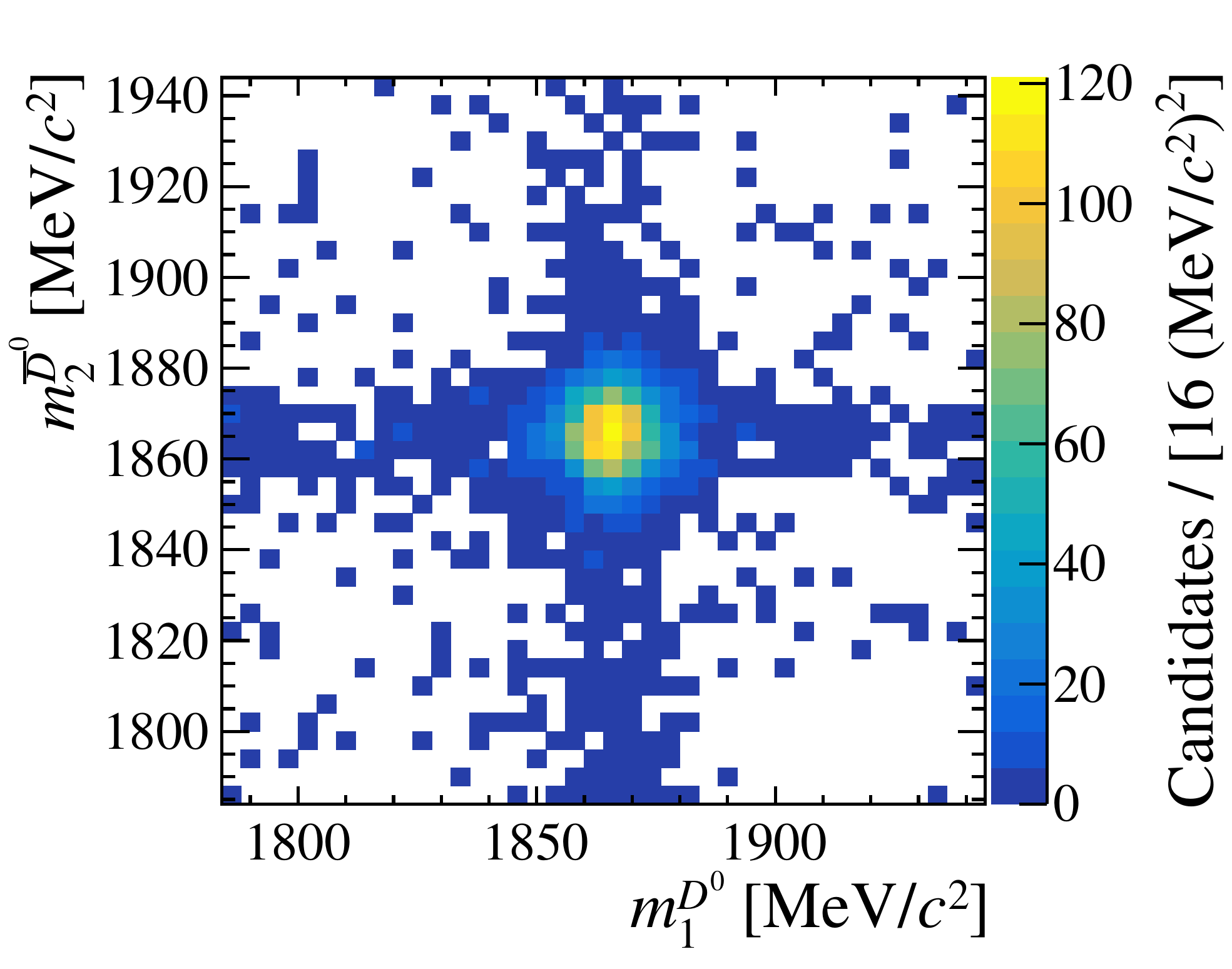}
  \includegraphics[width=0.66\floatwidth]{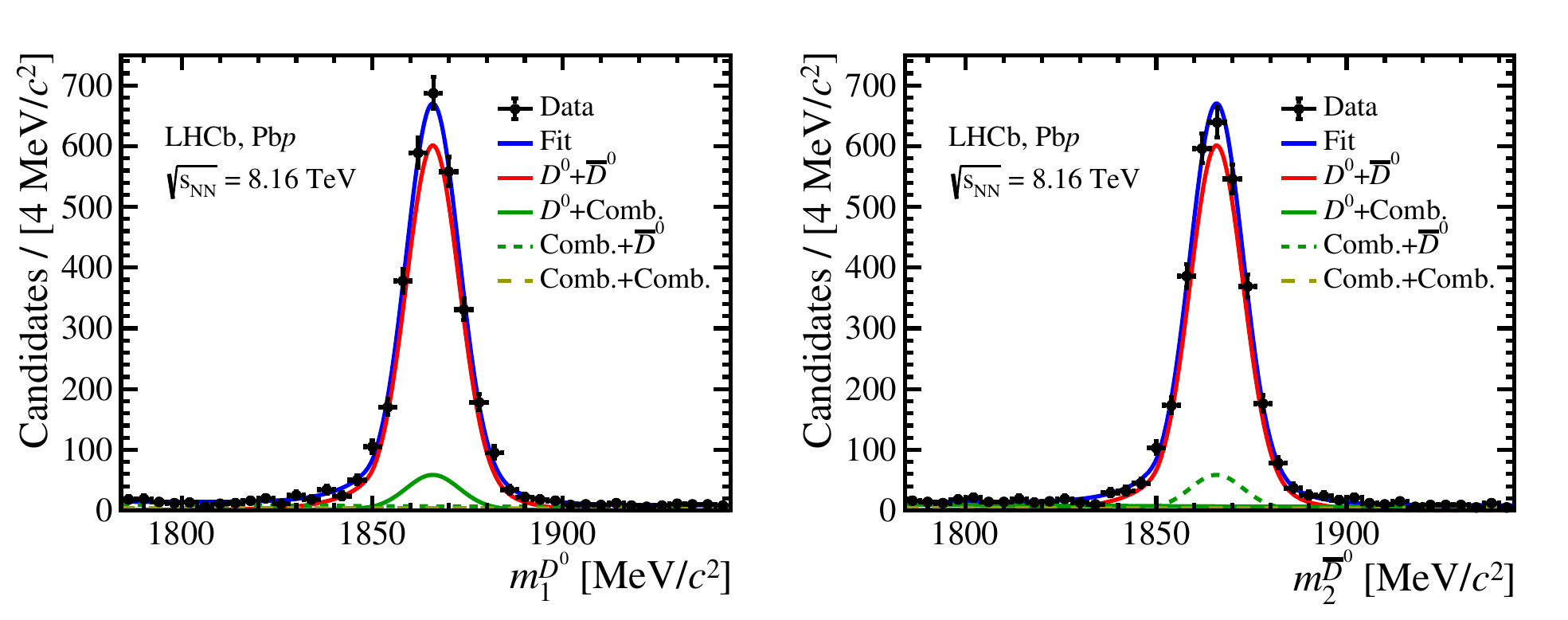}
  \caption{
  (Left) Two-dimensional invariant-mass distributions of $(m_1, m_2)$ for \pair{\Dz}{\Dzb} pairs and the projections on (middle) $m_1$ and (right) $m_2$ with the fit results superimposed. 
      Shown  in the projection plots are (points with error bars) data data, (solid blue) the total fit and the four fit  components. The plots in the top (bottom) row correspond to \pPb (\Pbp) data.
    }
  \label{fig:mass_DDb}
\end{figure}

\begin{figure}[!pbt]
\centering
  \includegraphics[width=0.33\floatwidth]{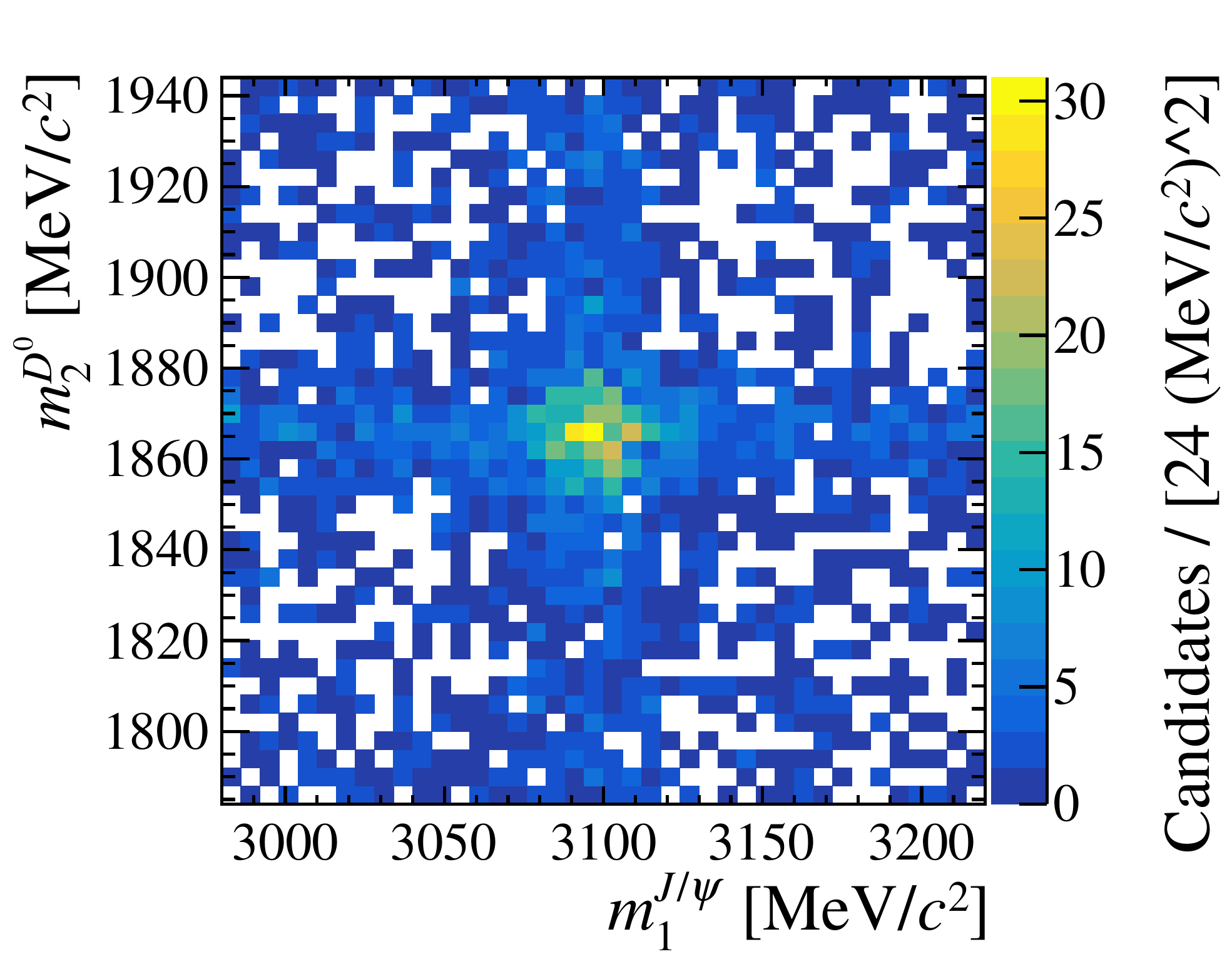}
  \includegraphics[width=0.66\floatwidth]{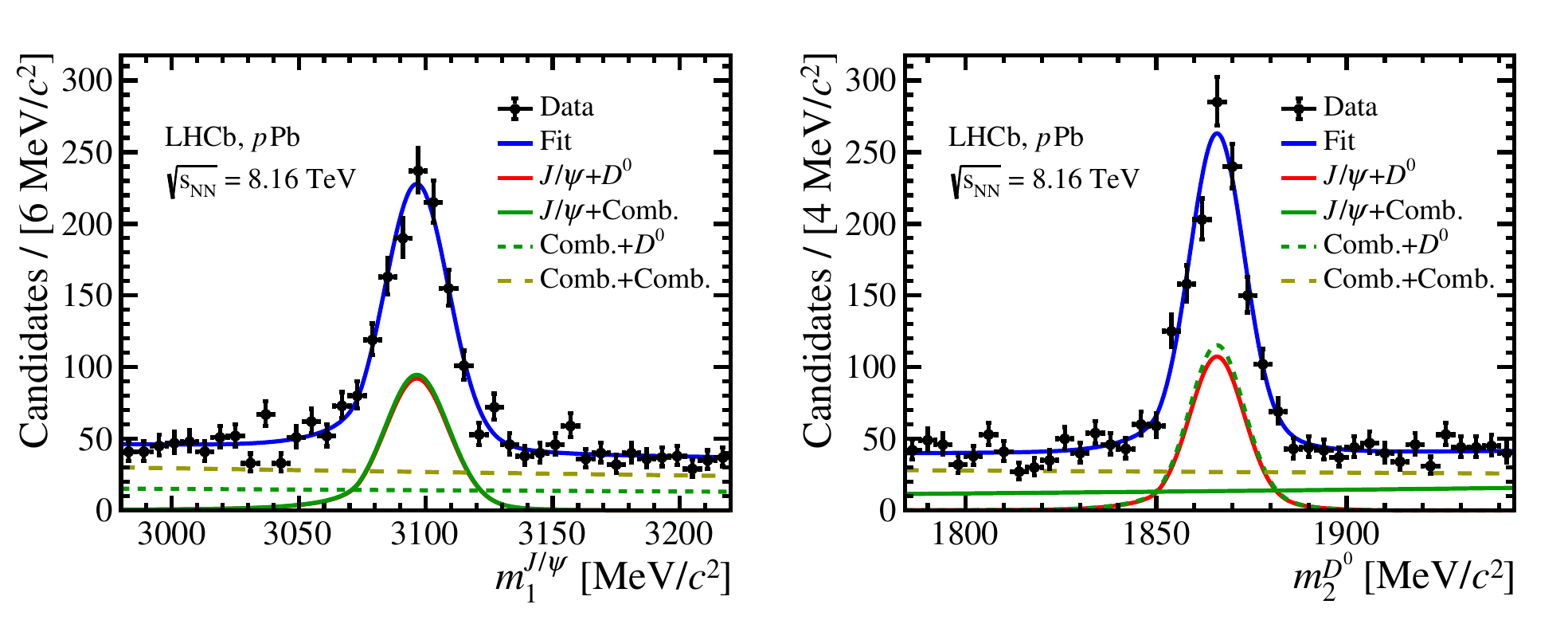}
  \includegraphics[width=0.33\floatwidth]{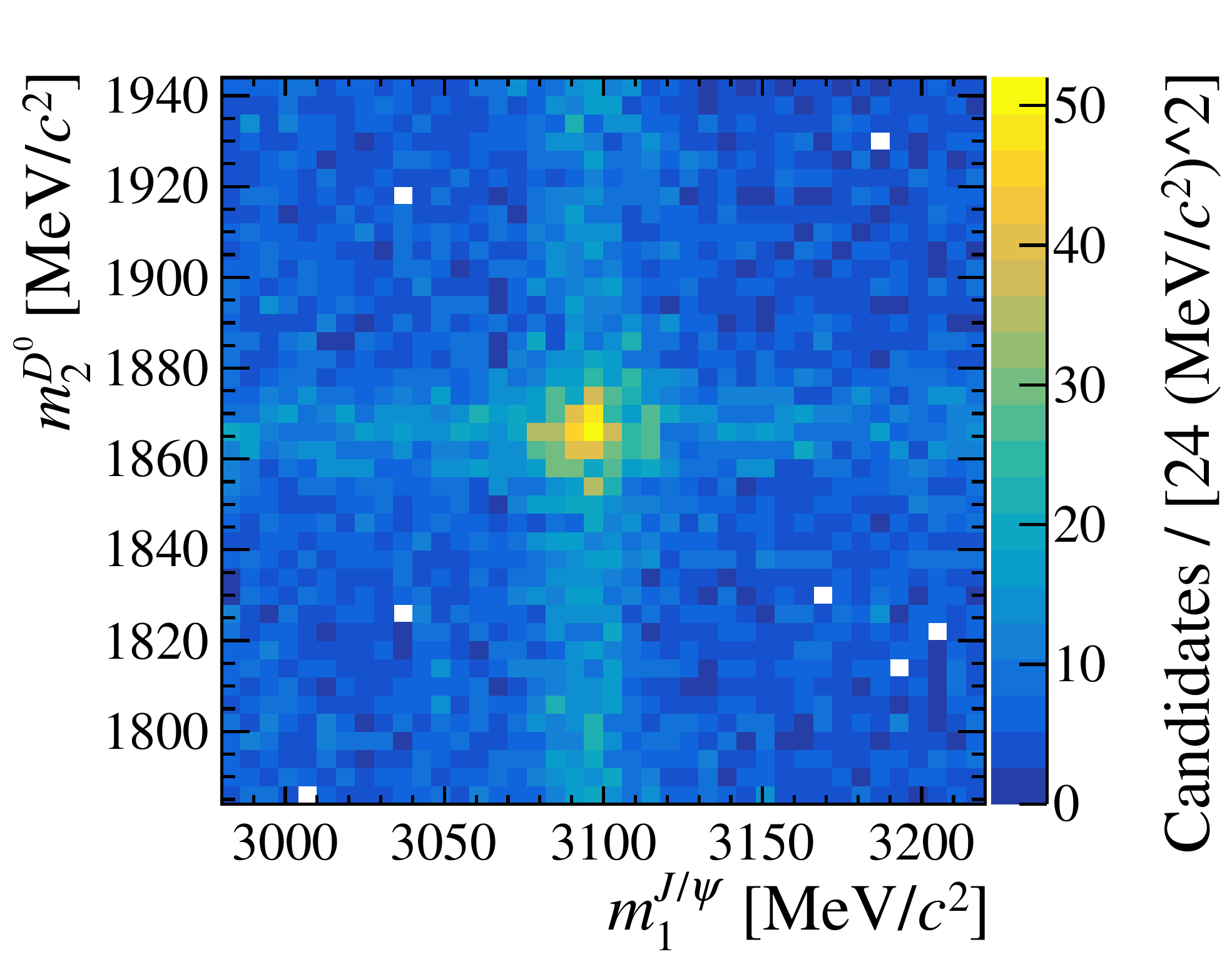}
  \includegraphics[width=0.66\floatwidth]{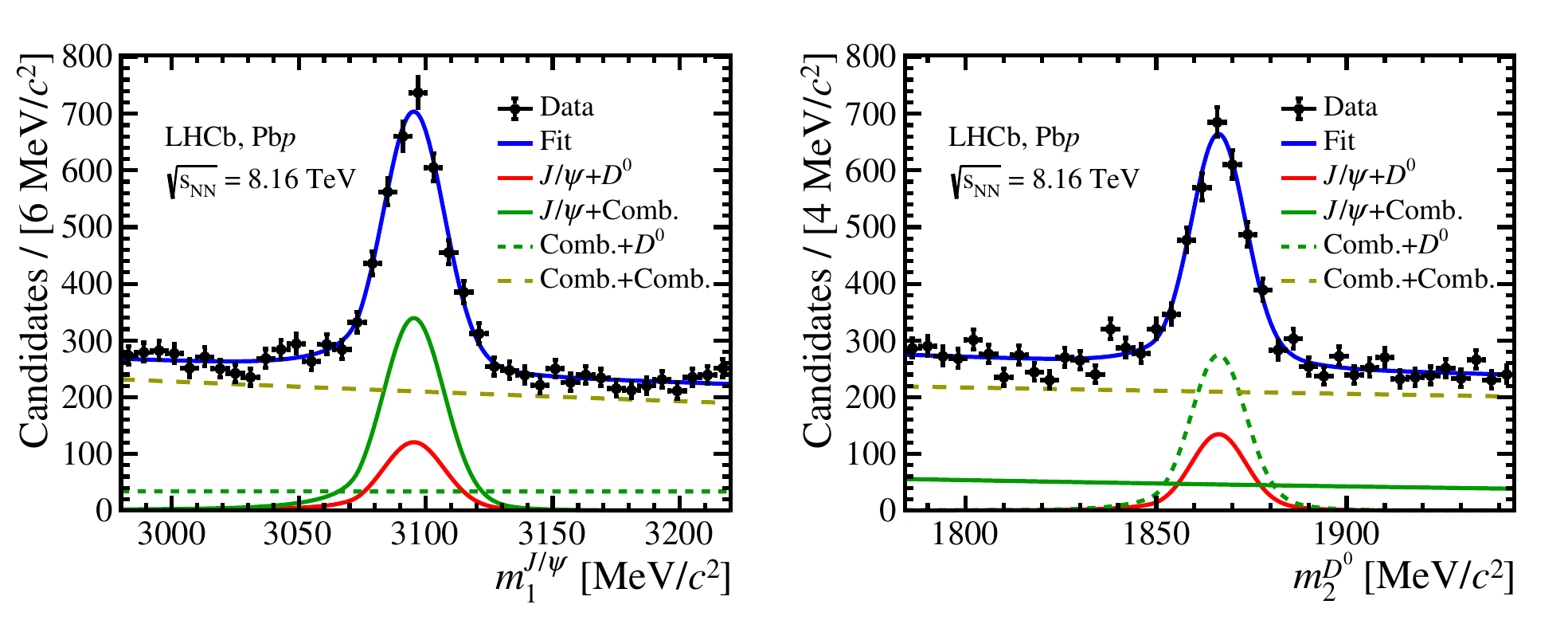}
  \caption{
  (Left) Two-dimensional invariant-mass distributions of $(m_1^\jpsi, m_2^\Dz)$ for \pair{\jpsi}{\Dz} pairs and the projections on (middle) $m_1^\jpsi$ and (right) $m_2^\Dz$ with the fit results superimposed. 
      Shown in the projection plots are (points with error bars) data data, (solid blue) the total fit and the four fit  components. The plots in the top (bottom) row correspond to \pPb (\Pbp) data.
    }
  \label{fig:mass_JpsiD}
\end{figure}

The invariant mass distributions and the fit projections are shown in Fig.~\ref{fig:mass_DD_Pbp} for \pair{\Dz}{\Dz} pairs in \Pbp data, and in Figs.~\ref{fig:mass_DDb} and  \ref{fig:mass_JpsiD} for \pair{\Dz}{\Dzb}  and \pair{\jpsi}{\Dz}  pairs in
both \pPb and \Pbp data.

\section*{Cross-sections and cross-section ratios}
The total cross-sections measured in the reduced charm hadron rapidity ($y(H_c)$) range $1.7<y(H_c)<3.7$ ($-4.7<y(H_c)<-2.7$) for \pPb (\Pbp) are  shown
in Table~\ref{tab:totCrossSection} under the requirement,  $\pt>0\gevc$ for $\jpsi, \Dz$ and $\pt>2\gevc$ for $\Dp,\Dsp$. The cross-section of
\pair{\jpsi}{\Dz} pair production is compared with the theoretical calculation of SPS production using the reweighted EPPS16 nuclear
PDF~\cite{shao:2015vga,Shao:2012iz,Kusina:2017gkz,Eskola:2016oht}.
 Those with the additional requirement $\pt(\Dz)>2\gevc$ are shown in Table~\ref{tab:totCrossSectionHpt}, and those in full rapidity acceptance for
\pair{\Dz}{\Dz}, \pair{\Dz}{\Dzb} and \pair{\jpsi}{\Dz} pairs are shown in Table~\ref{tab:totCrossSectionMore}.


\begin{table}[!bp]
    \caption{Total cross-sections of charm pair production (in \mbarn) in  \pPb and \Pbp data for the $\pt$ requirement on the charm hadron,  $\pt>0\gevc$ for $\jpsi, \Dz$ and $\pt>2\gevc$ for $\Dp,\Dsp$.
    The rapidity range is for each charm hadron in the pair.
    The first uncertainty is statistical and the second is systematic.
    The prediction of SPS \pair{\jpsi}{\Dz} production calculated using the weighted EPPS16 nuclear
    PDF~\cite{shao:2015vga,Shao:2012iz,Kusina:2017gkz,Eskola:2016oht} is also listed for comparison, where the first
    uncertainty is from uncertainties of scales, feed-down contribution and model parameters, and the second due to
    nuclear PDF uncertainties. 
    }
    \centering
    \begin{tabular}[h]{c|c|c}
Quantity & $1.7<y(H_c)<3.7$ & $-4.7<y(H_c)<-2.7$\\
\hline
       $\pair{\Dz}{\Dz }$& $11.0\pm0.7\pm1.2$ & $16.5\pm1.1\pm3.5$\\
       $\pair{\Dz}{\Dzb}$ & $34.1\pm1.2\pm3.6$ & $35.1\pm1.3\pm5.9$\\
       $\pair{\Dz}{\Dp }$ & $ 4.0\pm0.3\pm0.6$ & $ 5.3\pm0.4\pm1.7$\\
       $\pair{\Dz}{\Dm }$ & $12.7\pm0.4\pm1.7$ & $11.6\pm0.4\pm2.8$\\
       $\pair{\Dz}{\Dsp}$ & $ 2.1\pm0.6\pm0.3$ & $ 3.6\pm0.7\pm1.2$\\
       $\pair{\Dz}{\Dsm}$ & $ 5.4\pm0.6\pm0.8$ & $ 7.1\pm1.0\pm1.8$\\
       $\pair{\Dp}{\Dp }$ & $ 0.46\pm0.05\pm0.09$ & $ 0.53\pm0.08\pm0.24$\\
       $\pair{\Dp}{\Dm }$ & $ 1.38\pm0.07\pm0.22$ & $ 1.26\pm0.08\pm0.42$\\
       $\pair{\Dp}{\Dsm}$ & $ 1.43\pm0.21\pm0.25$ & $ 1.32\pm0.20\pm0.44$\\
       $\pair{\Dp}{\Dsp}$ & $ 0.32\pm0.12\pm0.05$ & $ 0.80\pm0.27\pm0.40$\\
        $\pair{\jpsi}{\Dz}$ & $0.46\pm0.05\pm0.04$ & $ 0.57\pm0.07\pm0.09$\\
       $\pair{\jpsi}{\Dp}$ & $0.08\pm0.01\pm0.01$ & $ 0.10\pm0.01\pm0.02$\\
       \hline
       \multirow{2}{*}{$\pair{\jpsi}{\Dz}$ SPS~\cite{shao:2015vga,Shao:2012iz,Kusina:2017gkz,Eskola:2016oht} } &
       \multirow{2}{*}{$0.051^{+0.467}_{-0.043} {}^{+0.009}_{-0.009}$}&
       \multirow{2}{*}{$0.055^{+0.426}_{-0.053}{}^{+0.004}_{-0.003}$}\\
        \rule{0pt}{1pt}&&\\
        \hline
    \end{tabular}
    \label{tab:totCrossSection}
\end{table}
\begin{table}[!bp]
    \caption{Total production cross-sections (in \mbarn) of open charm pairs involving the $\Dz$ meson
    in  \pPb and \Pbp data with the $\pt(\Dz)>2\gevc$ requirement. 
    The rapidity range is for each charm hadron in the pair.
    The first uncertainty is statistical and the second is systematic. Predictions for the $\Dz\,\Dz$ and $\Dz\,\Dzb$ cross-sections from Ref.~\cite{Helenius:2019uge} are given in the last two rows of the Table.
    }
    \centering
\begin{tabular}[h]{c|c|c}
Pairs & $1.7<y(H_c)<3.7$ & $-4.7<y(H_c)<-2.7$\\
\hline
    $\pair{\Dz}{\Dz}$ & $2.36\pm0.16\pm0.24$ & $2.28\pm0.15\pm0.33$\\
    $\pair{\Dz}{\Dzb}$ & $7.36\pm0.28\pm0.71$ & $6.01\pm0.24\pm0.81$\\
   $\pair{\Dz}{\Dp}$ & $1.83\pm0.12\pm0.26$ & $1.97\pm0.13\pm0.54$\\
   $\pair{\Dz}{\Dm}$ & $6.27\pm0.19\pm0.82$ & $5.06\pm0.17\pm1.07$\\
   $\pair{\Dz}{\Dsp}$ & $0.84\pm0.19\pm0.13$ & $1.46\pm0.50\pm0.38$\\
   $\pair{\Dz}{\Dsm}$ & $3.03\pm0.34\pm0.45$ & $2.38\pm0.29\pm0.50$\\
   $\pair{\jpsi}{\Dz}$ & $0.21\pm0.02\pm0.02$ & $0.22\pm0.02\pm0.03$\\
     \hline
     \multirow{2}{*}{$\pair{\Dz}{\Dz}$~\cite{Helenius:2019uge}} &\multirow{2}{*}{$3.62^{+5.92}_{-3.37}$} 
     &\multirow{2}{*}{$2.93^{+4.72}_{-2.69}$}\\
     \rule{0pt}{1pt}&&\\
 \multirow{2}{*}{$\pair{\Dz}{\Dzb}$ \cite{Helenius:2019uge}} 
 &  \multirow{2}{*}{$6.34^{+8.66}_{-4.79}$ } 
 & \multirow{2}{*}{$5.59^{+7.41}_{-3.79}$}\\
 \rule{0pt}{1pt}&&\\
        \hline
\end{tabular}
\label{tab:totCrossSectionHpt}
\end{table}
\begin{table}[!bp]
    \caption{Total cross-sections (in \mbarn) in  \pPb and \Pbp data for full rapidity acceptance without any $\pt(H_c)$ requirement.
    The first uncertainty is statistical and the second is systematic.
    }
    \centering
\begin{tabular}{c|c|c}
     Pairs& $1.5<y(H_c)<4$ & $-5<y(H_c)<-2.5$\\
     \hline
     $\pair{ \Dz}{\Dz}$ & $14.72\pm1.10\pm2.25$ & $24.05\pm2.12\pm5.19$\\
     $\pair{ \Dz}{\Dzb}$ & $45.99\pm2.09\pm5.04$ & $52.05\pm2.40\pm8.91$\\
     $\pair{\jpsi}{\Dz}$ & $\phantom{0}0.51\pm0.12\pm0.05$ & $\phantom{0}0.82\pm0.13\pm0.14$\\
     \hline
\end{tabular}
\label{tab:totCrossSectionMore}
\end{table}

\begin{figure}[!bp]
    \begin{center}
        \includegraphics[width=0.7\floatwidth]{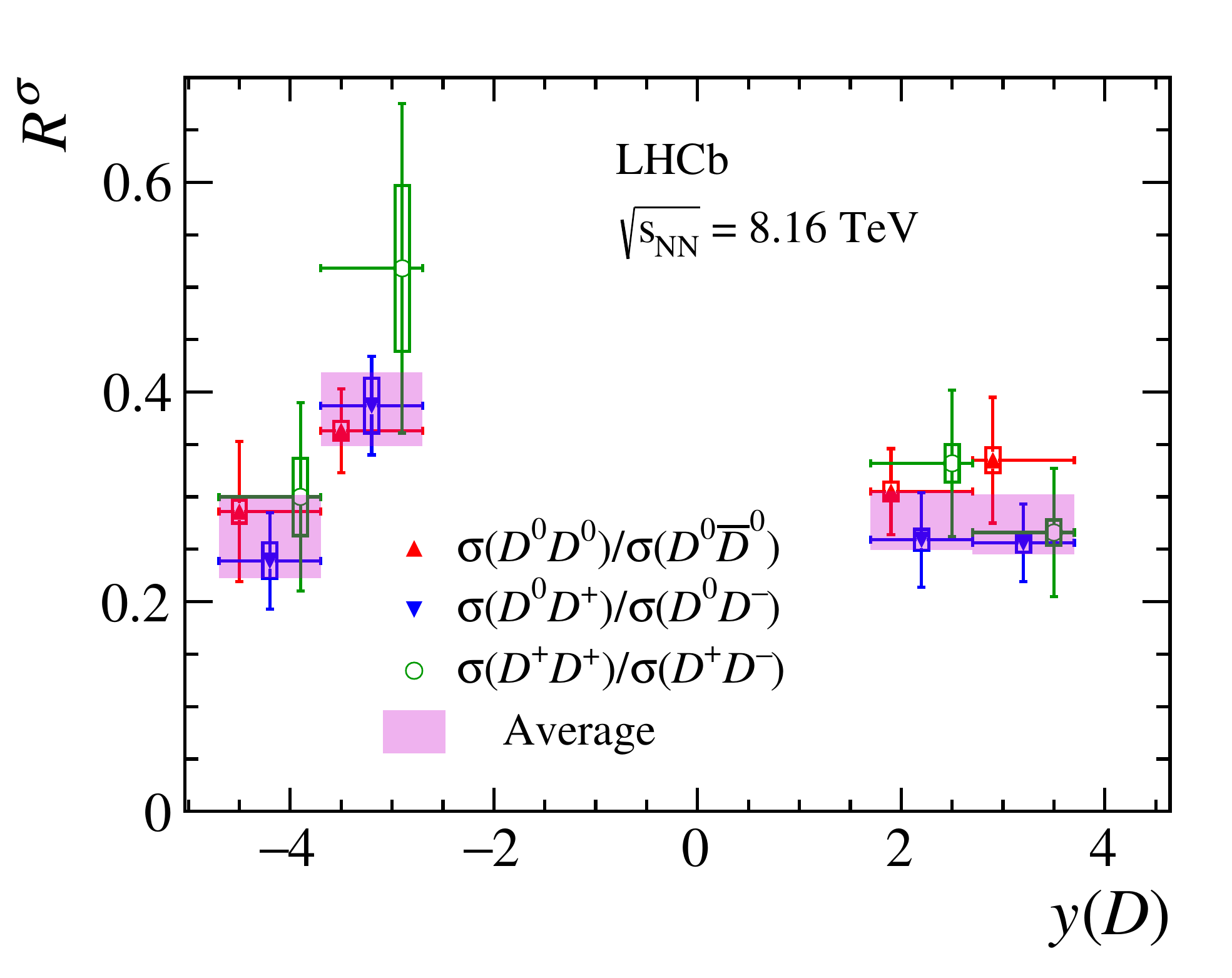}
    \end{center}
    \caption{
    Ratios of differential cross-sections in bins of charm hadron rapidity under the  $\pt(\D)>2\gevc$ requirement.
    The average over all the pairs listed in Table~\ref{tab:totCrossSectionRatioSSOS} for each rapidity interval is also presented.
    The boxes (bars) correspond to systematic (statistical) uncertainties. Points are shifted horizontally to improve visibility.
    }
    \label{fig:totCrossSectionRatioSSOS}
\end{figure}

In Table~\ref{tab:totCrossSectionRatioSSOS} and Fig.~\ref{fig:totCrossSectionRatioSSOS}, the ratio between cross-sections 
of like-sign and opposite-sign pairs is shown in bins of charm hadron rapidity for $\pt(H_c)>2\gevc$.


\begin{table}[!bp]
    \centering
    \caption{
        Ratios of differential cross-sections in bins of charm hadron rapidity for charm $\pt(H_c)>2\gevc$.
    The average over all the pairs in each rapidity interval is also presented.
    }
    \resizebox{\linewidth}{!}{
        \begin{tabular}{c|c|c|c|c}
Pairs &$-4.7<y(H_c)<-3.7$ & $-3.7<y(H_c)<-2.7$ & $1.7<y(H_c)<2.7$ & $2.7<y(H_c)<3.7$\\
\hline
            $\pair{\Dz}{\Dz}/  \pair{\Dz}{\Dzb}$& $0.286\pm0.067\pm0.011$ &$0.363\pm0.040\pm0.009$ &$0.305\pm0.041\pm0.009$ &$0.335\pm0.060\pm0.012$ \\
            $\pair{\Dz}{\Dp}/ \pair{\Dz}{\Dm}$& $0.239\pm0.046\pm0.017$ &$0.387\pm0.047\pm0.026$ &$0.259\pm0.045\pm0.010$ &$0.256\pm0.037\pm0.009$ \\
            $\pair{\Dp}{\Dp}/ \pair{\Dp}{\Dm}$& $0.300\pm0.090\pm0.037$ &$0.518\pm0.157\pm0.079$ &$0.332\pm0.070\pm0.018$ &$0.266\pm0.061\pm0.012$ \\
            $\pair{\Dz}{\Dsp}/\pair{\Dz}{\Dsm}$& $0.321\pm0.260\pm0.043$ &$0.504\pm0.172\pm0.028$ &$0.201\pm0.096\pm0.006$ &$0.303\pm0.139\pm0.014$ \\
            $\pair{\Dp}{\Dsp}/\pair{\Dp}{\Dsm}$& $0.291\pm0.590\pm0.080$ &$0.582\pm0.340\pm0.113$ &$0.143\pm0.116\pm0.009$ &$0.234\pm0.113\pm0.013$ \\
            \hline
Average&$0.262\pm0.035\pm0.019$&$0.384\pm0.029\pm0.019$&$0.278\pm0.026\pm0.010$&$0.274\pm0.027\pm0.011$\\
            \hline
        \end{tabular}
    }
    \label{tab:totCrossSectionRatioSSOS}
\end{table}

\newpage
\addcontentsline{toc}{section}{References}
\bibliographystyle{LHCb}
\bibliography{main,standard,LHCb-PAPER,LHCb-CONF,LHCb-DP,LHCb-TDR}

\newpage
\centerline
{\large\bf LHCb collaboration}
\begin
{flushleft}
\small
R.~Aaij$^{31}$,
C.~Abell{\'a}n~Beteta$^{49}$,
T.~Ackernley$^{59}$,
B.~Adeva$^{45}$,
M.~Adinolfi$^{53}$,
H.~Afsharnia$^{9}$,
C.A.~Aidala$^{83}$,
S.~Aiola$^{25}$,
Z.~Ajaltouni$^{9}$,
S.~Akar$^{64}$
% ...

{\footnotesize \it

% ...
$ ^{9}$Universit{\'e} Clermont Auvergne, CNRS/IN2P3, LPC, Clermont-Ferrand, France\\
% ...
$ ^{25}$INFN Sezione di Milano, Milano, Italy\\
% ...
$ ^{31}$Nikhef National Institute for Subatomic Physics, Amsterdam, Netherlands\\
% ...
$ ^{45}$Instituto Galego de F{\'\i}sica de Altas Enerx{\'\i}as (IGFAE), Universidade de Santiago de Compostela, Santiago de Compostela, Spain\\
% ...
$ ^{49}$Physik-Institut, Universit{\"a}t Z{\"u}rich, Z{\"u}rich, Switzerland\\
% ...
$ ^{53}$H.H. Wills Physics Laboratory, University of Bristol, Bristol, United Kingdom\\
% ...
$ ^{59}$Oliver Lodge Laboratory, University of Liverpool, Liverpool, United Kingdom\\
% ...
$ ^{64}$University of Cincinnati, Cincinnati, OH, United States\\
% ...
$ ^{67}$Syracuse University, Syracuse, NY, United States\\
% ...
$ ^{83}$University of Michigan, Ann Arbor, United States, associated to $^{67}$\\
\bigskip
% ...
\medskip
}
\end{flushleft}

\end{document}